\documentclass[12pt]{iopart}
\usepackage{iopams}
\usepackage{amssymb}
\usepackage{graphicx,subfigure,epstopdf}

\usepackage{cite}
\eqnobysec
\usepackage[colorlinks,linktocpage,linkcolor=blue]{hyperref}
\begin{document}

\title[]{Localization and ballistic diffusion for the tempered fractional Brownian-Langevin motion}

\author{Yao Chen, Xudong Wang, and Weihua Deng}

\address{School of Mathematics and Statistics, Gansu Key Laboratory of Applied Mathematics and Complex Systems, Lanzhou University, Lanzhou 730000, P.R. China}
\ead{ychen2015@lzu.edu.cn, xdwang14@lzu.edu.cn, and dengwh@lzu.edu.cn}
\vspace{10pt}

\begin{abstract}
This paper further discusses the tempered fractional Brownian motion, its ergodicity, and the derivation of the corresponding Fokker-Planck equation. Then we introduce the generalized Langevin equation with the tempered fractional Gaussian noise for a free particle, called tempered fractional Langevin equation (tfLe). While the tempered fractional Brownian motion displays localization diffusion for the long time limit and for the short time its mean squared displacement has the asymptotic form $t^{2H}$, 
 we show that the asymptotic form of the mean squared displacement of the tfLe transits from $t^2$ (ballistic diffusion for short time) to $t^{2-2H}$, and then to $t^2$ (again ballistic diffusion for long time). On the other hand, the overdamped tfLe has the  transition of the diffusion type from $t^{2-2H}$ to $t^2$ (ballistic diffusion). The tfLe with harmonic potential is also considered.
%

\noindent (Some figures may appear in colour only in the online journal)
\end{abstract}

\section{Introduction}


Revealing the mechanism of the motion of particles in complex disordered systems is a fundamental and challenging topic \cite{Shlesinger,Hughes}. Generally, the motion is no longer Brownian because of the complex environment and/or the properties of the particles themselves. The mean squared displacement (MSD) of Brownian motion goes like $\langle(\Delta x)^2\rangle=\langle[x(t)-\langle x(t)\rangle]^2\rangle \sim t^\nu$ with $\nu=1$; for a long time $t$ if $\nu \not= 1$, it is called anomalous diffusion, being subdiffusion for $\nu <1$ and superdiffusion for $\nu > 1$ \cite{Metzler}; in particular, it is termed as localization diffusion if $\nu=0$ and ballistic diffusion if $\nu=2$. There are two types of stochastic processes to model anomalous diffusion: Gaussian processes and non-Gaussian ones; the non-Gaussian processes \cite{Drysdale,Metzler:01,Schumer:03,Bruno:00} include continuous time random walks (CTRWs), L\'evy processes, subordinated L\'evy processes, and the Gaussian ones \cite{Kou,Pipiras,Deng,Meerschaert:02,Eric} contain fractional Brownian motion (fBm), generalized Langevin equation, etc. In this paper, we mainly consider the Langevin type equation with correlated internal noise, being a Gaussian process.

The most basic Gaussian process, describing normal diffusion, is Brownian motion with its corresponding Langevin equation \cite{Coffey:04}
\begin{eqnarray}
m\frac{\rmd v}{\rmd t}=-\xi v+F(t),\nonumber
\end{eqnarray}
where $v$ is the velocity of a Brownian particle with a mass $m$, $\xi $ is a frictional constant, and the random fluctuation force $F(t)$ is white noise. For large $t$, the mean squared displacement of a Brownian particle is $\langle (\Delta x(t))^2\rangle=\langle(\int_0^t v(s)\rmd s)^2\rangle\simeq 2Dt$ with $D=\frac{k_B T}{\xi}$ being the Einstein relation, where $k_B$ is the Boltzmann constant and $T$ the absolute temperature of the environment. As the extension of Brownian motion, fractional Brownian motion (fBm) is still a Gaussian process, which can be seen as the fractional derivative (or integral) of a Brownian motion. The fBm is defined as \cite{Mandelbrot:68}
\begin{eqnarray}
B_H(t)=\frac{1}{\Gamma(1-\alpha)}\int_{-\infty}^{+\infty}[(t-s)_+^{-\alpha}-(-s)_+^{-\alpha}]B(\rmd s),\nonumber
\end{eqnarray}
where
\begin{equation*}
\label{cases}
(x)_+=\cases{x&for $x > 0$\\
0&for $x \leq 0$\\}
\end{equation*}
with $-\frac{1}{2}<\alpha<\frac{1}{2}$, and the Hurst index $H=\frac{1}{2}-\alpha$. Note that Brownian motion is recovered when $H=\frac{1}{2}$. The variance of $ B_H(t)$ is $2D_Ht^{2H}$, where $D_H=[\Gamma(1-2H)\cos(H\pi)]/(2H \pi)$.
Fractional Langevin equation \cite{Deng,Eric}, still being Gaussian process, reads
\begin{eqnarray}
m\frac{\rmd ^2x(t)}{\rmd t^2}=-\xi \int_0^t(t-\tau)^{2H-2}\frac{\rmd x}{\rmd\tau}\rmd\tau+\eta \gamma(t),\nonumber
\end{eqnarray}
where $x(t)$ is the particle displacement, $\eta=[k_BT\xi/(2D_HH(2H-1))]^{1/2}$, $\gamma(t)=\rmd B_H(t)/\rmd t$ is the fractional Gaussian noise, and $1/2<H<1$ is the Hurst parameter. 
The mean squared displacement of the trajectory sample $x(t)$ for large $t$ is $\langle x^2(t)\rangle\simeq2k_BT/(\xi\Gamma(2H-1)\Gamma(3-2H))t^{2-2H}$. 

Along the direction of extension of Brownian motion, the new Gaussian process, called tempered fractional Brownian motion (tfBm) \cite{Meerschaert:00}, was introduced by Meerschaert and Sabzikar with its definition
\begin{eqnarray}\label{PDF}
B_{\alpha,\lambda}(t)=\int_{-\infty}^{+\infty}[\rme^{-\lambda(t-x)_+}(t-x)_+^{-\alpha}-\rme^{-\lambda(-x)_+}(-x)_+^{-\alpha}]B(\rmd x), \end{eqnarray}
where $\lambda>0$, $\alpha<\frac{1}{2}$, and the Hurst index $H=\frac{1}{2}-\alpha$;
and its basic theory was developed with application to modeling wind speed. This paper naturally introduces the generalized Langevin equation, termed as tempered fractional Langevin equation (tfLe), with the tempered fractional noise as internal noise. We discuss the ergodicity of the tfBm, and derive the corresponding Fokker-Planck equation. The mean squared displacement (MSD) of the tfLe is carefully analyzed, displaying the transition from $t^2$ (ballistic diffusion for short time) to $t^{2-2H}$, and then to $t^2$ (again ballistic diffusion for long time). For the overdamped tfLe, its MSD transits from $t^{2-2H}$ to $t^2$ (ballistic diffusion). The properties of the correlation function of the tfLe with harmonic potential is also considered.

The outline of this paper is as follows. In Section \ref{two}, we review the tfBm, derive its Fokker-Planck equation, and discuss its ergodicity. The tfLe is introduced in Section \ref{three}, in which the underdamped case, overdamped case, and the tfLe with harmonic potential are discussed in detail. We conclude the paper with some discussions in the last section.

\section{Tempered fractional Brownian motion and tempered fractional Gaussian noise}\label{two}
\setcounter{equation}{0}
We introduce the definitions of the tfBm and the tempered fractional Gaussian noise (tfGn). The Fokker-Planck equation for tfBm is derived, and the ergodicity of the tfBm is discussed.
\subsection{Tempered fractional Brownian motion}
Tempered fractional Brownian motion is defined in (\ref{PDF}), which modifies the power law kernel in the moving average representation of a fractional Brownian motion by adding an exponential tempering \cite{Meerschaert:00}; it is generalizedly self-similar in the sense that for any $c>0$
\begin{equation} \label{self-similar}
\{ B_{\alpha,\lambda}(ct) \}_{t \in \mathbb{R}} = \{ c^{H} B_{\alpha,c\lambda}(t) \}_{t \in \mathbb{R}} 
\end{equation}
in distribution and it has the covariance function
%
%

 \begin{eqnarray}\label{covariance}
 \textrm{Cov}[B_{\alpha,\lambda}(t),B_{\alpha,\lambda}(s)]=\frac{1}{2}\left[C_t^2|t|^{2H}+C_s^2|s|^{2H}-C_{t-s}^2|t-s|^{2H}\right]
 \end{eqnarray}
 for any $t,s\in\mathbb{R}$, where 
 \begin{eqnarray} \label{Cdelta}
 C_t^2&=\int_{-\infty}^{+\infty}\left[\rme^{-\lambda|t|(1-x)_+}(1-x)_+^{-\alpha}-\rme^{-\lambda|t|(-x)_+}(-x)_+^{-\alpha}\right]^2\rmd x\\ &=\frac{2\Gamma(2H)}{(2\lambda|t|)^{2H}}-\frac{2\Gamma(H+\frac{1}{2})K_H(\lambda|t|)}{\sqrt{\pi}(2\lambda|t|)^H}\nonumber
 \end{eqnarray}
 for $t\neq 0$ and $C_0^2=0$, where $K_H(x)$ is the modified Bessel function of the second kind \cite{Meerschaert:00}. It is obvious that the variance of tfBm is $\langle B_{\alpha,\lambda}^2(t)\rangle=C_t^2|t|^{2H}\simeq 2\Gamma(2H)(2\lambda)^{-2H}$ as $t\rightarrow \infty$ on account of $K_H( t)\simeq\sqrt{\pi}(2 t)^{-\frac{1}{2}}\rme ^{ -t}$, which means that tfBm is localization. And $\langle B_{\alpha,\lambda}(t)\rangle=0$. Since $B_{\alpha,\lambda}(t)$ is a Gaussian process, when $t>0$, $B_{\alpha,\lambda}(t)\sim N(0,C_t^2t^{2H})$. Then it has the probability density function (PDF)
\begin{eqnarray}\label{PDB}
P(x,t)=\frac{1}{\sqrt{2\pi C_t^2t^{2H}}}\rme^{-\frac{x^2}{2C_t^2t^{2H}}}.
\end{eqnarray}
The Fourier transform of (\ref{PDB}) is $P(k,t)=\int_{-\infty}^{+\infty}\rme ^{\rmi kx}P(x,t)\rmd x=\rme ^{-C_t^2t^{2H}k^2/2}$; taking partial derivative w.r.t. $t$ and performing inverse Fourier transform on both sides of this equation lead to
\begin{eqnarray}\label{PDA}
\frac{\partial P(x,t)}{\partial t}=-\frac{\Gamma(H+\frac{1}{2})}{\sqrt{\pi}(2\lambda)^H}\left[Ht^{H-1}K_H(\lambda t)+t^H\dot K_H(\lambda t)\right]\frac{\partial^2P(x,t)}{\partial x^2}.
\end{eqnarray}
For large $t$, from (\ref{PDA}), we have that the PDF is asymptotically governed by
\begin{eqnarray}
\frac{\partial P(x,t)}{\partial t}=\frac{\Gamma(H+\frac{1}{2})\rme^{-\lambda t}}{(2\lambda)^{H+\frac{1}{2}}}\left[\lambda t^{H-\frac{1}{2}}-\left(H-\frac{1}{2}\right)t^{H-\frac{3}{2}}\right]\frac{\partial^2P(x,t)}{\partial x^2}.
\end{eqnarray}
The non-Markovian property is implied by the time-dependent diffusion constant: $D_{\alpha,\lambda}(t)=(2\lambda)^{-H-1/2}\Gamma(H+1/2)\rme^{-\lambda t}[\lambda t^{H-1/2}-(H-1/2)t^{H-3/2}]$.

\begin{figure}
  \flushright
\includegraphics[scale=0.4]{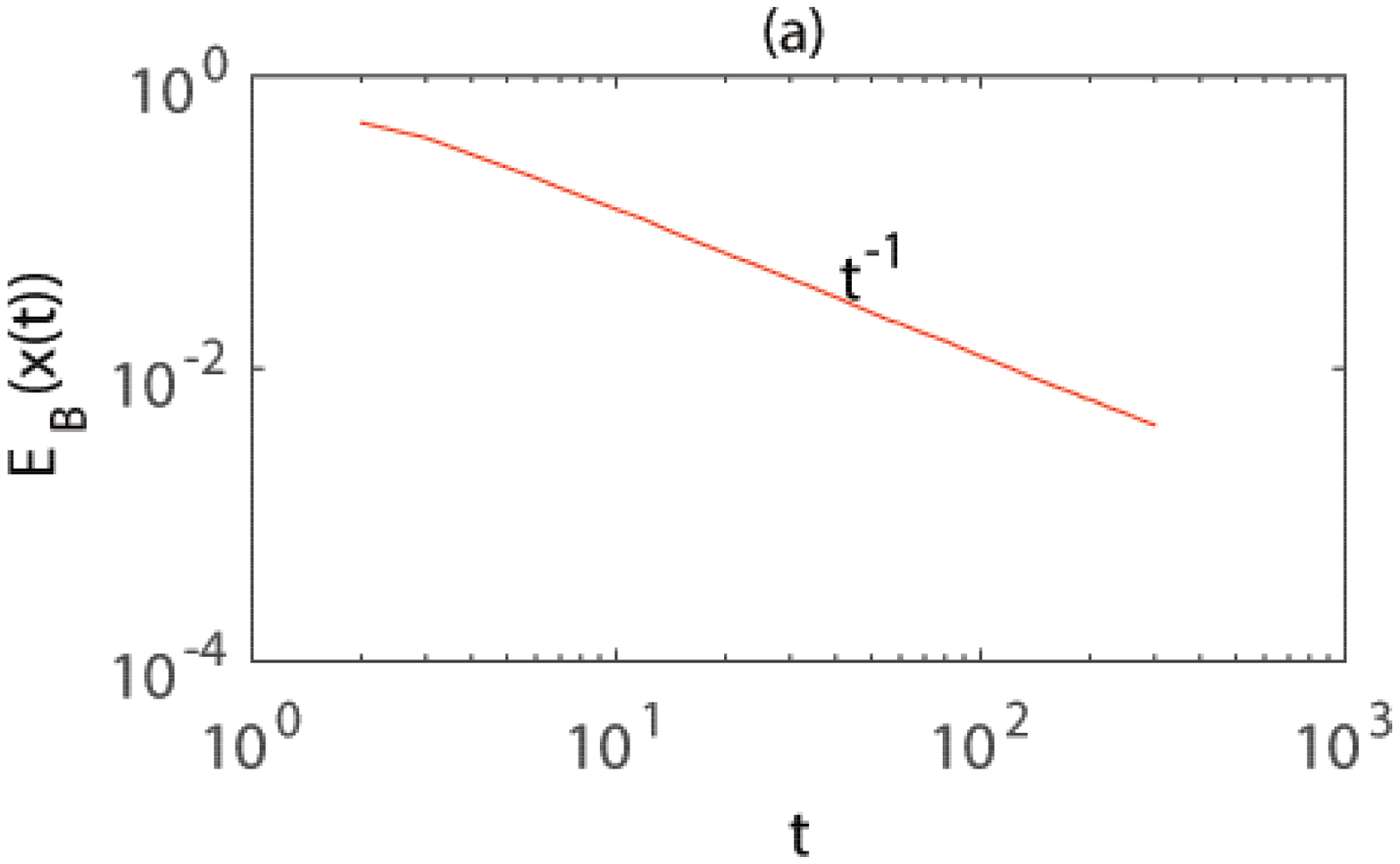}\includegraphics[scale=0.4]{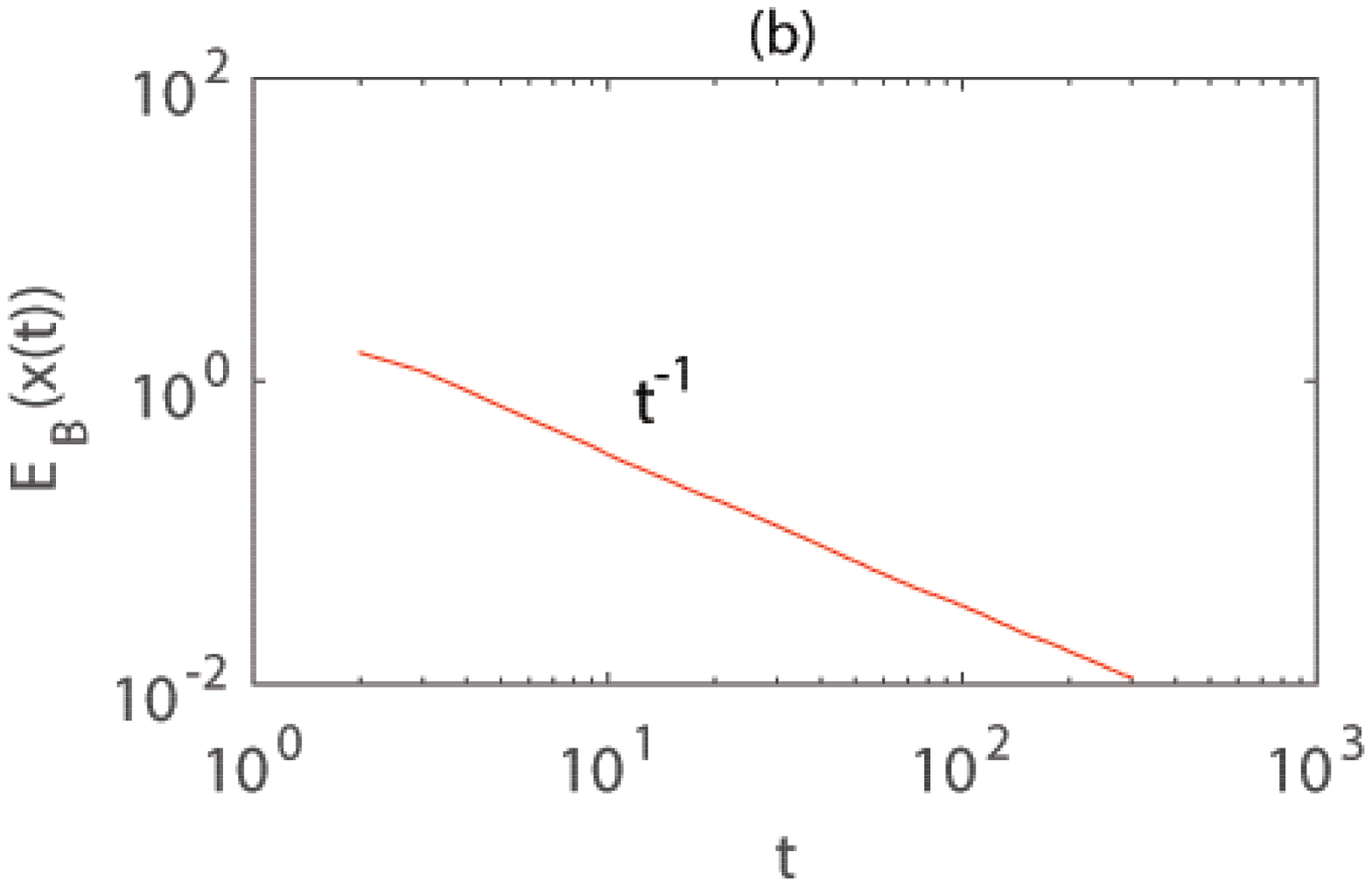}
  \caption{Solid (red) lines are the simulation results for $E_B(x(t))$. The parameters $H$, $\lambda$, $\Delta$ and $T$ are, respectively, taken as $H=0.8$ (left panel (a)), $H=0.4$ (right panel (b)), $\lambda=0.1$, $\Delta=1$ and $T=300$. The 5000 trajectories are sampled. 
  }
  \label{fig:0}
\end{figure}

Next we consider the ergodicity of tfBm and the convergence speed of the variance of the time-average mean-squared displacement, $\bar{\delta}^2(B_{\alpha,\lambda}(t))=\int_0^{t-\Delta}[B_{\alpha,\lambda}(t'+\Delta)-B_{\alpha,\lambda}(t')]^2\rmd t'/(t-\Delta)$, where $\Delta$ is the lag time. If the average of $\bar{\delta}^2(B_{\alpha,\lambda}(t))$ equals to the ensemble average of $B_{\alpha,\lambda}(t)$ and the variance of $\bar{\delta}^2(B_{\alpha,\lambda}(t))$ tends to zero when the measurement time is long, the process tfBm is ergodic \cite{Deng}; it is indeed (see (\ref{qi}) and (\ref{fang}) and their derivations presented in \ref{AppendixA}). In fact,
the variance of $\bar{\delta}^2(B_{\alpha,\lambda}(t))$ is a measure of ergodicity breaking and the ergodicity breaking parameter is defined as
\begin{eqnarray}
E_B(B_{\alpha,\lambda}(t))=\frac{\textrm{Var}[\bar{\delta}^2(B_{\alpha,\lambda}(t))]}{\langle \bar{\delta}^2(B_{\alpha,\lambda}(t))\rangle^2}=\frac{\langle [\bar{\delta}^2(B_{\alpha,\lambda}(t))]^2\rangle-\langle \bar{\delta}^2(B_{\alpha,\lambda}(t))\rangle^2}{\langle \bar{\delta}^2(B_{\alpha,\lambda}(t))\rangle^2}.\nonumber
\end{eqnarray}
For tfBm, from \ref{AppendixA} the average of $\bar{\delta}^2(B_{\alpha,\lambda}(t))$ is 
\begin{eqnarray}\label{qi}
\langle\bar{\delta}^2(B_{\alpha,\lambda}(t))\rangle
=C_ {\Delta}^2|\Delta|^{2H},
\end{eqnarray}
hence $\langle\bar{\delta}^2\rangle=\langle B_{\alpha,\lambda}^2\rangle$ for all times; for long times and moderate $\lambda$, the variance of $\bar{\delta}^2(B_{\alpha,\lambda}(t))$ is
\begin{eqnarray}\label{fang}
\textrm{Var}[\bar{\delta}^2(B_{\alpha,\lambda}(t))]
\simeq D\frac{4\Gamma^2(H+\frac{1}{2})}{(2\lambda)^{2H+1}}\Delta^{2H}t^{-1},
\end{eqnarray}
where 
$D=\int_0^{\infty}\rmd\tau[(1+\tau)^{H-\frac{1}{2}}\rme^{-\lambda\Delta(1+\tau)}+|\tau-1|^{H-\frac{1}{2}}\rme^{-\lambda|\Delta(\tau-1)|}-2\tau^{H-\frac{1}{2}}\rme^{-\lambda\Delta\tau}]^2$.  
Thus we have $E_B(B_{\alpha,\lambda}(t))\simeq 4D\Gamma^2(H+1/2)/[(2\lambda)^{2H+1}C_{\Delta}^4\Delta^{2H}]t^{-1}$, which tends to zero with the speed $t^{-1}$ for $H \in (0,1)$ as  $t\rightarrow\infty$. 
The simulation result for ergodicity breaking parameter $E_B(B_{\alpha,\lambda}(t))$ is shown in Figure \ref{fig:0}.
On the other hand, for the limit $\lambda t\rightarrow0$, $C_t^2$ tends to a constant. In this case, (\ref{covariance}) reduces to the covariance function of fractional Brownian motion (fBm). Then the evolution of the ergodicity breaking parameter $E_B(B_{\alpha,\lambda}(t))$ is recovered to that of fBm \cite{Deng}.

\subsection{Tempered fractional Gaussian noise}
Given a tfBm (\ref{PDF}), we can define the tempered fractional Gaussian noise (tfGn)
\begin{eqnarray}\label{2.9}
\gamma(t)=\frac{B_{\alpha,\lambda}(t+h)-B_{\alpha,\lambda}(t)}{h},
\end{eqnarray}
being similar to the definition of fractional Gaussian noise (fGn) \cite{Mandelbrot:68}, where $h$ is small and $h\ll t$. 

Based on the covariance function (\ref{covariance}) of $B_{\alpha,\lambda}(t)$ and its zero mean, one can easily obtain that the mean of tfGn is $\langle \gamma(t)\rangle=0$ and its covariance is
\begin{eqnarray*}
\fl \langle\gamma(t_1)\gamma(t_2)\rangle
=\frac{1}{h^2}\langle(B_{\alpha,\lambda}(t_1+h)-B_{\alpha,\lambda}(t_1))(B_{\alpha,\lambda}(t_2+h)-B_{\alpha,\lambda}(t_2))\rangle \nonumber \\
\fl\qquad\quad\qquad=\frac{1}{2h^2}(C_{t_1-t_2+h}^2|t_1-t_2+h|^{2H}+C_{t_1-t_2-h}^2|t_1-t_2-h|^{2H}-2C_{t_1-t_2}^2|t_1-t_2|^{2H}),
\end{eqnarray*}
which means that the tfGn is a stationary Gaussian process.  
 For a fixed $\lambda$, the asymptotic behavior of the covariance function is $\langle\gamma(0)\gamma(t)\rangle\simeq-\Gamma(H+1/2)\lambda^2(2\lambda)^{-H-1/2}t^{H-1/2}\rme^{-\lambda t}$ for large $ t$ and $\langle\gamma(0)\gamma(t)\rangle\simeq C_t^2H(2H-1)t^{2H-2}$ with a positive constant $C_t^2$ for small $ t$ (for the details of derivation, see \ref{AppendixB}).

\section{Generalized Langevin equation with tempered fractional Gaussian noise}\label{three}
\setcounter{equation}{0}

We discuss the dynamics of the generalized Langevin equation with tempered fractional Gaussian noise for free particles, including the underdamped and overdamped cases. The properties of the correlation function of the equation under external potential are also discussed.

\subsection{Dynamical behaviors for free particles}

\subsubsection{Underdamped generalized Langevin equation}

We know that for large $t$ the MSD of fBm is like $ t^{2H}$ and the one of the corresponding Langevin equation is like $ t^{2-2H}$. In the above section, it is shown that tfBm is localization, i.e., its MSD goes like $t^0$. Can we expect that the MSD of the tfLe goes like $t^2$ for large $t$? The answer is yes.

Based on the second fluctuation-dissipation theorem \cite{Kubo}, which links the dissipation memory kernel $K(t)$ with the autocorrelation function of internal noise $F(t)$: $\langle F(t_1)F(t_2)\rangle=k_BT\xi K(t_1-t_2)$, the tfLe can be written as
\begin{eqnarray}\label{PDC}
m\frac{\rmd ^2x(t)}{\rmd t^2}=-\xi \int_0^tK(t-\tau)\frac{\rmd x}{\rmd\tau}\rmd\tau+F(t),
\end{eqnarray}
with $\dot x(0)=v_0$, $x(0)=0$, where $v_0$ is the initial velocity, $F(t)=\sqrt{2k_BT\xi}\gamma(t)$ is the internal noise with tfGn $\gamma(t)$, and $K(t)=2\langle\gamma(0)\gamma(t)\rangle=\frac{1}{h^2}(C_{t+h}^2|t+h|^{2H}+C_{t-h}^2|t-h|^{2H}-2C_t^2|t|^{2H})$ with  $0<H<1$.
In this case, the fluctuation and dissipation stem from the same source and the system will finally reach the equilibrium state. 
Taking Laplace transform of (\ref{PDC}) leads to
\begin{eqnarray} \label{3.2}
x(s)=\frac{F(s)+mv_0}{ms^2+\xi s K(s)}.
\end{eqnarray}

For convenience, denoting $C_t^2t^{2H}$ by $f(t)$, then $K(s)=\mathcal{L}[K(t)]=\frac{1}{h^2}(\rme^{sh}f(s)+\rme^{-sh}f(s)-2f(s))$. No matter $s\rightarrow0$ or $s\rightarrow\infty$, $sh$ is always small since $h\ll t$. So along with a Taylor's series expansion, we have $K(s)\simeq s^2f(s),\nonumber$ for $s\rightarrow0$ or $s\rightarrow\infty$. That is to say, the asymptotic expression of $x(s)$ is
\begin{eqnarray}\label{PDW}
x(s)\simeq\frac{F(s)+mv_0}{ms^2+\xi s^3 f(s)}
\end{eqnarray}
and
\begin{eqnarray}\label{PDWX}
v(s)\simeq\frac{F(s)+mv_0}{ms+\xi s^2 f(s)}.
\end{eqnarray}
Based on (\ref{PDW}) and (\ref{PDWX}), we do the dynamics analysis for different cases.

\begin{figure}
	\flushright
	\includegraphics[scale=0.37]{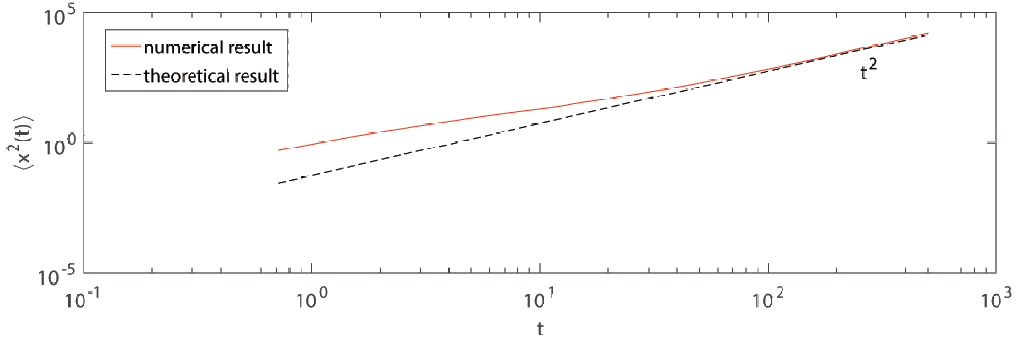}\\
	\caption{Theoretical result (\ref{xx}) (black dashed line) and computer simulation one sampled over 1000 trajectories (red solid line), plotted for $H=0.7$, $\lambda=0.1$, $T=500$, $k_BT=1$, $\xi=1$, $m=1$, and $v_0=1$. 
	}\label{fig:1}
\end{figure}

\textit{Case I} ($t$ is large, i.e., $s \rightarrow 0$,  and $H \in (0,1)$): For moderate $\lambda$, $\lambda t$ is large, say, $\lambda t> 10$. Then  $f(t)=C_t^2t^{2H}\simeq 2\Gamma(2H)(2\lambda)^{-2H}$.
 By final value theorem, we have $\lim\limits_{s\to0}f(s)=2\Gamma(2H)(2\lambda)^{-2H}/s$;  substituting it into (\ref{PDW}) leads to
\begin{eqnarray}
x(s)\simeq\frac{F(s)+mv_0}{As^2}
\end{eqnarray}
with $A=m+2\xi \Gamma(2H)(2\lambda)^{-2H}$. 
Then
\begin{eqnarray}
x(t)\simeq\frac{1}{A}\int_0^t\tau F(t-\tau)\rmd\tau+\frac{mv_0}{A}t.
\end{eqnarray}
Hence $\langle x(t)\rangle\simeq\frac{mv_0}{A}t$ for large $t$, and
\begin{eqnarray}
\langle x^2(t)\rangle\simeq\frac{2}{A^2}\int_0^t\int_0^{t_1}\rmd t_2\rmd t_1t_1t_2\langle F(t-t_1)F(t-t_2)\rangle+\frac{m^2v_0^2}{A^2}t^2,
\end{eqnarray}
with
$\langle F(t-t_1)F(t-t_2)\rangle=k_BT\xi h^{-2}
(C_{t_1-t_2+h}^2|t_1-t_2+h|^{2H}+C_{t_1-t_2-h}^2|t_1-t_2-h|^{2H}-2C_{t_1-t_2}^2|t_1-t_2|^{2H})$.
 Let $g(t_1)=\int_0^{t_1}t_2\langle F(t-t_1)F(t-t_2)\rangle \rmd t_2$. Using the property of convolution and Taylor's series expansion, we have $g(s)=\mathcal{L}[g(t_1)]\simeq k_BT\xi f(s)$. So $g(t_1)\simeq k_BT\xi f(t_1)$ for large $t_1$. Then
\begin{eqnarray}\label{xx}
\langle x^2(t)\rangle&\simeq\frac{2}{A^2}\int_0^tt_1g(t_1)\rmd t_1+\frac{m^2v_0^2}{A^2}t^2 \nonumber\\
&\simeq\frac{2k_BT\xi}{A^2}\int_0^tt_1C_{t_1}^2t_1^{2H}\rmd t_1+\frac{m^2v_0^2}{A^2}t^2 \nonumber\\
&\simeq\frac{2k_BT\xi\Gamma(2H)+m^2v_0^2(2\lambda)^{2H}}{A^2(2\lambda)^{2H}}t^2
\end{eqnarray}
for $t\rightarrow\infty$. Eq. (\ref{xx}) is confirmed by the numerical simulations; see Figure \ref{fig:1}. And the MSD of the system is
\begin{eqnarray}
\langle(\Delta x)^2\rangle=\langle[x(t)-\langle x(t)\rangle]^2\rangle
\simeq\frac{2k_BT\xi\Gamma(2H)}{A^2(2\lambda)^{2H}}t^2,
\end{eqnarray}
displaying the ballistic diffusion.

Besides, let us further consider the autocorrelation function of the position $x(t)$. After making double Laplace transforms of the autocorrelation function of internal noise $\langle F(t_1)F(t_2)\rangle$, we have
\begin{eqnarray}
\langle F(s_1)F(s_2)\rangle=k_BT\xi\frac{s_1^2f(s_1)+s_2^2f(s_2)}{s_1+s_2}.
\end{eqnarray}
Then
\begin{eqnarray}\label{ii}
\langle x(s_1)x(s_2)\rangle&=\frac{\langle F(s_1)F(s_2)\rangle+m^2v_0^2}{A^2s_1^2s_2^2}\nonumber\\
&=\frac{k_BT\xi}{A^2}\frac{s_1^2f(s_1)+s_2^2f(s_2)}{(s_1+s_2)s_1^2s_2^2}+\frac{1}{A^2}\frac{m^2v_0^2}{s_1^2s_2^2}.
\end{eqnarray}
Taking inverse Laplace transforms of (\ref{ii}) and letting $t_1, t_2$ tend to infinity, we obtain
\begin{eqnarray}
\langle x(t_1)x(t_2)\rangle \simeq\left(\frac{2k_BT\xi\Gamma(2H)}{A^2(2\lambda)^{2H}}+\frac{m^2v_0^2}{A^2}\right) t_1t_2,
\end{eqnarray}
which results in (\ref{xx}) when $t_1=t_2$.

Furthermore, if $\lambda$ is moderate and $t\rightarrow\infty$, the velocity 
\begin{eqnarray}
v(t)&\simeq\frac{1}{A}\int_0^t F(\tau)\rmd\tau+\frac{mv_0}{A}
=\frac{\sqrt{2k_BT\xi}}{A}B_{\alpha,\lambda}(t)+\frac{mv_0}{A}.
\end{eqnarray}
Then we have $\langle v(t)\rangle\simeq\frac{mv_0}{A}$ and the autocorrelation function of $v(t)$ is
\begin{eqnarray}\label{vv}
\langle v(t_1)v(t_2)\rangle&\simeq\frac{2k_BT\xi}{A^2}\langle B_{\alpha,\lambda}(t_1)B_{\alpha,\lambda}(t_2)\rangle+\frac{m^2v_0^2}{A^2}\nonumber\\
&=\frac{k_BT\xi}{A^2}(C_{t_1}^2t_1^{2H}+C_{t_2}^2t_2^{2H}-C_{t_1-t_2}^2|t_1-t_2|^{2H})+\frac{m^2v_0^2}{A^2}.
\end{eqnarray}
If $t_1=t_2=t$, then
\begin{eqnarray}
\langle v^2(t)\rangle
&=\frac{2k_BT\xi}{A^2}C_t^2t^{2H}+\frac{m^2v_0^2}{A^2}\simeq \frac{4k_BT\xi\Gamma(2H)+m^2v_0^2(2\lambda)^{2H}}{A^2(2\lambda)^{2H}}
\end{eqnarray}
as $t\rightarrow\infty$. From (\ref{vv}), the second moment of position $x(t)$ is
\begin{eqnarray}
\fl\langle x^2(t)\rangle
=2\int_0^t\int_0^{t_1}\rmd t_2\rmd t_1\langle v(t_1)v(t_2)\rangle\nonumber\\
\fl\qquad\quad=\frac{2k_BT\xi}{A^2}\int_0^t\int_0^{t_1}\rmd t_2\rmd t_1(C_{t_1}^2{t_1}^{2H}+C_{t_2}^2{t_2}^{2H}-C_{t_1-t_2}^2|t_1-t_2|^{2H})+\frac{m^2v_0^2}{A^2}t^2\nonumber\\
\fl\qquad\quad\simeq\frac{2k_BT\xi\Gamma(2H)+m^2v_0^2(2\lambda)^{2H}}{A^2(2\lambda)^{2H}}t^2,
\end{eqnarray}
where $C_t^2t^{2H}\simeq 2\Gamma(2H)(2\lambda)^{-2H}$ for large $\lambda t$ is used; Eq. (\ref{xx}) is confirmed again. The correlation function of $x(t)$ and $v(t)$ is
\begin{eqnarray}
\langle x(t)v(t)\rangle&=\frac{1}{2}\frac{\rmd\langle x^2(t)\rangle}{\rmd t}\simeq\frac{2k_BT\xi\Gamma(2H)+m^2v_0^2(2\lambda)^{2H}}{A^2(2\lambda)^{2H}}t.
\end{eqnarray}

\textit{Remark}: There is another way to obtain the asymptotic expression of $f(s)$ for small $s$. By using the formula \cite{Srivastava}:
 \begin{eqnarray*}
 I_{\rho}^{\mu}(s)&=\int_0^{\infty}\rme^{-st}t^{\mu-1}K_{\nu}(\alpha t^{\rho})\rmd t\nonumber\\
 &=\frac{2^{(\mu-2\rho)/\rho}}{\rho \alpha^{\mu/\rho}}\sum\frac{(-s)^n}{n!}\Big(\frac{2}{\alpha}\Big)^{n/\rho}\Gamma\Big(\frac{\mu+n}{2\rho-\frac{1}{2}\nu}\Big)\Gamma\Big(\frac{\mu+n}{2\rho+\frac{1}{2}\nu}\Big),
\end{eqnarray*}
 where $\rho>0$, $\rm{Re}(s)>0$, $ \rm{Re}(\alpha)>0$, $\rm{Re}(\mu/\rho)>|\rm{Re}(\nu)|$ and $K_{\nu}(\alpha t)$ is the modified Bessel function of the second kind, we obtain the Laplace transform of  $f(t)\,(=C_t^2t^{2H})$ as
\begin{eqnarray*}
\fl f(s)=\frac{2\Gamma(2H)(2\lambda)^{-2H}}{s}-\frac{\Gamma(H+\frac{1}{2})}{\sqrt{\pi}\lambda^{2H+1}}\sum\frac{(-s)^n}{n!}\Big(\frac{2}{\lambda}\Big)^n\Gamma\Big(\frac{H+1+n}{2-\frac{1}{2}H}\Big)\Gamma\Big(\frac{H+1+n}{2+\frac{1}{2}H}\Big).
\end{eqnarray*}
So $f(s) \simeq\frac{2\Gamma(2H)(2\lambda)^{-2H}}{s}$ for small $s$.

\begin{figure}
  \flushright
  \includegraphics[scale=0.37]{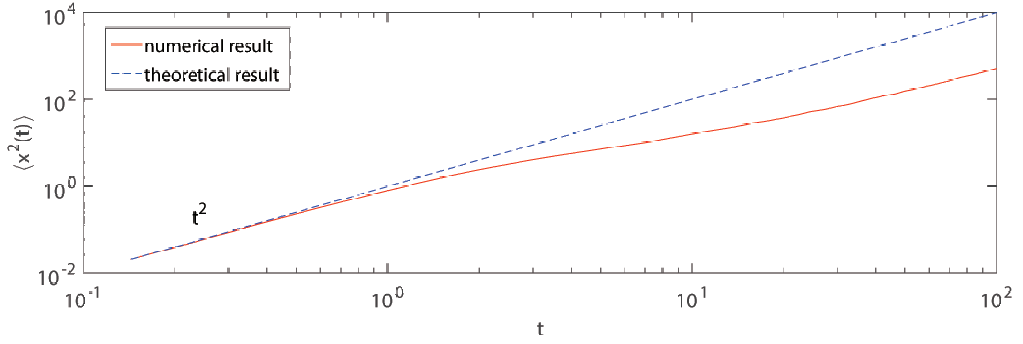}\\
  \caption{Theoretical result (\ref{xiao}) (blue dashed line) and numerical one sampled over $1000$ trajectories (red solid line), plotted for $H=0.8$, $\lambda=0.1$, $T=100$, and $v_0=1$. }\label{2}
\end{figure}

\textit{Case II} ($s$ is large, i.e., $t \rightarrow 0$, and $H \in (\frac{1}{2},1)$): 
For moderate or small $\lambda$, $\lambda t\rightarrow0$. Then we have $C_t^2
\simeq 2D_H\Gamma^2(H+1/2)$ and  $f(t)\simeq 2D_H\Gamma^2(H+1/2)t^{2H}$ with its Laplace transform  $f(s) \simeq 2D_H\Gamma^2(H+1/2)\Gamma(2H+1)s^{-1-2H}$. So (\ref{PDW}) becomes
\begin{eqnarray} \label{3.18}
x(s)\simeq\frac{F(s)+mv_0}{ms^2+ 2\xi D_H\Gamma^2(H+\frac{1}{2})\Gamma(2H+1)s^{2-2H} }\simeq\frac{F(s)+mv_0}{ms^2}
\end{eqnarray}
for large $s$. Taking inverse Laplace transform on Eq. (\ref{3.18}) leads to
\begin{eqnarray}
x(t)\simeq\frac{1}{m}\int_0^t\tau F(t-\tau)\rmd\tau+v_0t.
\end{eqnarray}
Then the mean $\langle x(t)\rangle\simeq v_0t$. Similarly to the case $t\rightarrow \infty$, we obtain
\begin{eqnarray}\label{xiao}
\langle x^2(t)\rangle&\simeq\frac{2}{m^2}\int_0^t\int_0^{\tau_1}\rmd\tau_2\rmd\tau_1\tau_1\tau_2\langle F(t-\tau_1)F(t-\tau_2)\rangle+v_0^2t^2\nonumber \\
&=\frac{8D_H\Gamma^2(H+\frac{1}{2})k_BT\xi}{m^2}\int_0^t\tau_1^{2H+1}\rmd\tau_1+v_0^2t^2\nonumber\\
&=\frac{4D_H\Gamma^2(H+\frac{1}{2})k_BT\xi}{m^2(H+1)}t^{2H+2}+v_0^2t^2\nonumber\\
&\simeq v_0^2t^2
\end{eqnarray}
for small $t$. Note that for short times we have $\langle x^2(t)\rangle\simeq\frac{k_BT}{m}t^2$ if the thermal initial condition $\langle v_0^2\rangle=\frac{k_BT}{m}$ is assumed. Figure \ref{2} numerically confirms the theoretical result. Naturally, in this case, $v(t)\simeq \frac{1}{m}\int_0^tF(\tau)\rmd\tau+v_0$. So $\langle v(t)\rangle\simeq v_0$ and $\langle v^2(t)\rangle\simeq\frac{4k_BT\xi D_H\Gamma^2(H+\frac{1}{2})}{m^2}t^{2H}+v_0^2\simeq v_0^2$ for small $t$.

\textit{Case III} ($\lambda$ is small, $t$ is large, and $H \in (\frac{1}{2},1)$): 
If $\lambda t\rightarrow\infty$, it reduces to \textit{Case I}. Now we only consider the case that $\lambda t$ is small. As the second case, when $\lambda t$ is small, $C_t^2\simeq 2D_H\Gamma^2(H+1/2)$. Then (\ref{PDW}) becomes
\begin{eqnarray}
x(s)&\simeq\frac{F(s)+mv_0}{ms^2+ 2\xi D_H\Gamma^2(H+\frac{1}{2})\Gamma(2H+1)s^{2-2H} }\nonumber\\
&\simeq\frac{F(s)+mv_0}{2\xi D_H\Gamma^2(H+\frac{1}{2})\Gamma(2H+1)s^{2-2H}}
\end{eqnarray}
for small $s$, which has the inverse Laplace transform 
\begin{eqnarray}
x(t)\simeq P\int_0^tF(t-\tau)\tau^{1-2H}\rmd\tau+mv_0 P t^{1-2H}
\end{eqnarray}
with $P=\frac{1}{2\xi D_H\Gamma^2(H+\frac{1}{2})\Gamma(2H+1)\Gamma(2-2H)}$. So finally we obtain $\langle x(t)\rangle\simeq mv_0 P t^{1-2H}$ and the MSD is
\begin{eqnarray}\label{zhong}
\langle x^2(t)\rangle\simeq\frac{k_BT}{\xi D_H\Gamma^2(H+\frac{1}{2})\Gamma(2H+1)\Gamma(3-2H)}t^{2-2H}.
\end{eqnarray}
Similarly, $v(t)\simeq\frac{1}{m}\int_0^tF(\tau)E_{2H}(-B(t-\tau)^{2H})\rmd\tau+v_0E_{2H}(-Bt^{2H})$ by making the inverse Laplace transform of $v(s)\simeq\frac{F(s)+mv_0}{ms+ 2\xi D_H\Gamma^2(H+\frac{1}{2})\Gamma(2H+1)s^{1-2H} }$, where $B=2\xi D_H\Gamma^2(H+\frac{1}{2})\Gamma(2H+1)m^{-1}$ and $E_{\alpha}(z)$ is the particular case of the generalized Mittag-Leffler function \cite{Podlubny}
\begin{eqnarray}
E_{\alpha,\beta}(z)=\sum_{k=0}^{\infty}\frac{z^k}{\Gamma(\alpha k+\beta)},\qquad\alpha>0,\qquad \beta>0 \nonumber
\end{eqnarray}
with $E_{\alpha}(z)=E_{\alpha,1}(z)$. The asymptotic expression of the generalized Mittag-Leffler function for $z\rightarrow\infty$ is $E_{\alpha,\beta}(-z)\simeq\frac{z^{-1}}{\Gamma(\beta-\alpha)}$, where $0<\alpha<1$ or $1<\alpha<2$.

It can be noted that $\langle F(t_1)F(t_2)\rangle\simeq4k_BT\xi D_H\Gamma^2(H+1/2)H(2H-1)|t_1-t_2|^{2H-2}$ when $\lambda t$ is small, being similar to the autocorrelation function of fGn \cite{Deng}; and the expression of $v(t)$ is similar to the one of fLe. So in this case, the asymptotic behavior of $\langle v(t)\rangle$ and $\langle v^2(t)\rangle$ is consistent with that of fLe. Thus $\langle v(t)\rangle\simeq v_0E_{2H}(-Bt^{2H})\simeq t^{-2H}$ and $\langle v^2(t)\rangle$ decays like $t^{-4H}$ for large $t$ \cite{Eric}.

\begin{figure}
  \flushright
  \includegraphics[scale=0.37]{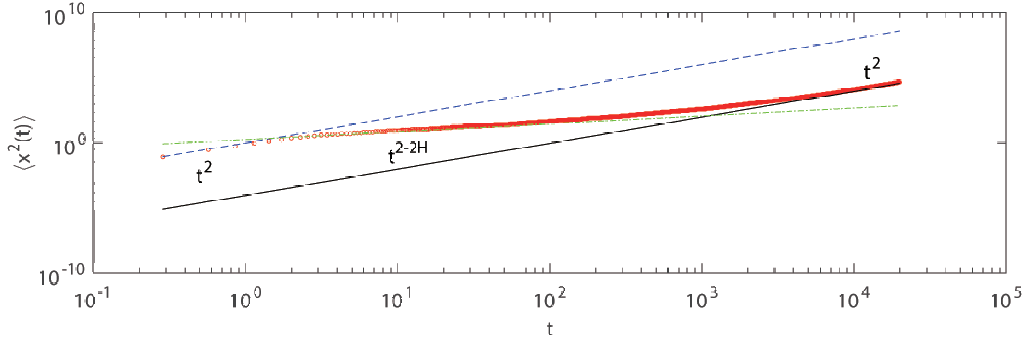}\\
  \caption{Theoretical results (\ref{xiao}) (blue dashed line),  (\ref{zhong}) (green dotted-dashed line), (\ref{xx}) (black solid line), and numerical result sampled over $1000$ trajectories (red data points), plotted for $H=0.7$, $\lambda=0.001$, $T=2\times 10^4$, $k_BT=1$, $\xi=1$, $m=1$, $v_0=1$, and $D_H=\frac{1}{2}$.
  }\label{3}
\end{figure}

From the above discussions, we know that the MSD of tfLe strongly depends on the value of $\lambda$ and $t$. When $\lambda t$ is small, it displays the same asymptotic behavior as the one of fLe \cite{Deng}. As time goes on, for fixed $\lambda$ the MSD of tfLe transmits as $t^2\to t^{2-2H}\to t^2$; the smaller $\lambda$ is, the longer the time of the MSD behaves as $t^{2-2H}$. See Figure \ref{3} for the demonstration of the results.


\subsubsection{Overdamped generalized Langevin equation}
\begin{figure}
  \flushright
  \includegraphics[scale=0.37]{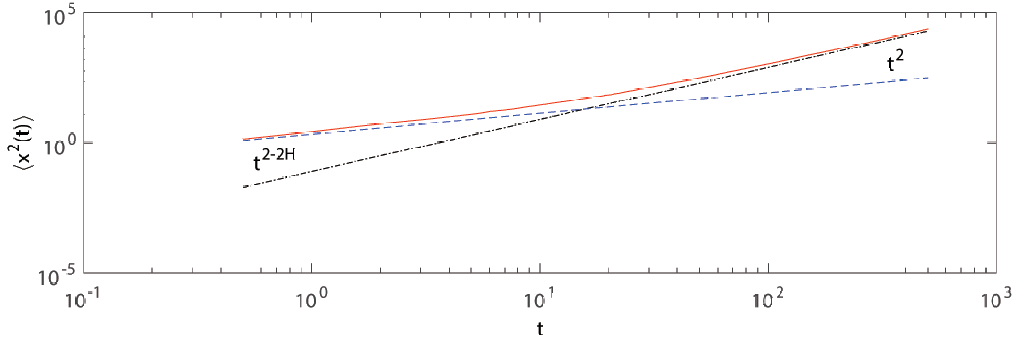}\\
  \caption{Theoretical results (\ref{over2}) (blue dashed line) and (\ref{over1}) (black dotted-dashed line), and numerical result sampled over $1000$ trajectories (red solid line), plotted for $H=0.6$, $\lambda=0.1$, $T=500$, $k_BT=1$, $\xi=1$, and $D_H=\frac{1}{2}$.
  }\label{3-1}
\end{figure}

The overdamped generalized Langevin equation without  Newton's acceleration term reads as
\begin{eqnarray}
0=-\xi \int_0^tK(t-\tau)\frac{\rmd x}{\rmd\tau}\rmd\tau+F(t),
\end{eqnarray}
where $K(t)$ and $F(t)$ are the same as the ones in Eq.  (\ref{PDC}). For moderate $\lambda$, taking the same procedure as above subsection, we can easily obtain that $\langle x(t)\rangle=0$ and 
\begin{eqnarray}\label{over1}
\langle x^2(t)\rangle\simeq\frac{2k_BT\xi\Gamma(2H)}{(2\xi\Gamma(2H)(2\lambda)^{-2H})^2(2\lambda)^{2H}}t^2
\end{eqnarray}
for large $t$, being the same as the underdamped case (\ref{PDC}). For short times,
\begin{eqnarray}
x(t)\simeq\frac{1}{2D_H\Gamma^2(H+\frac{1}{2})\xi\Gamma(2H+1)\Gamma(2-2H)}\int_0^tF(t-\tau)\tau^{1-2H}\rmd\tau;
\end{eqnarray}
then we have $\langle x(t)\rangle=0$ and the MSD
\begin{eqnarray}\label{over2}
\langle x^2(t)\rangle\simeq\frac{k_BT}{\xi D_H\Gamma^2(H+\frac{1}{2})\Gamma(2H+1)\Gamma(3-2H)}t^{2-2H}.
\end{eqnarray}
It can be noted that for short times, the MSD of the overdamped tfLe transits from $t^{2-2H}$ to $t^2$, and it behaves the same as the one of overdamped fLe \cite{Deng} for small $\lambda t$. See Figure \ref{3-1} for the numerical simulations, which verify (\ref{over1}) and (\ref{over2}).

\subsection{Harmonic potential}
Now we further consider the tfLe (\ref{PDC}) with external potential $U(x)$, i.e.,
\begin{eqnarray}
m\frac{\rmd^2x(t)}{\rmd t^2}=-\xi \int_0^tK(t-\tau)\frac{\rmd x}{\rmd\tau}\rmd\tau-U'(x)+F(t),
\end{eqnarray}
where $-U'(x)$ is an external force. If the external potential is a harmonic potential $U(x)=\frac{1}{2}m\omega^2x^2(t)$, where $\omega$ is the frequency of the oscillator, then we have
\begin{eqnarray}\label{ex}
m\frac{\rmd^2x(t)}{\rmd t^2}=-\xi \int_0^tK(t-\tau)\frac{\rmd x}{\rmd\tau}\rmd\tau-m\omega^2x(t)+F(t).
\end{eqnarray}
In what follows we analyze the normalized displacement correlation function, which is defined by
\begin{eqnarray}
C_x(t)=\frac{\langle x(t)x(0)\rangle}{\langle x^2(0)\rangle},\nonumber
\end{eqnarray}
under the thermal initial conditions $\langle F(t)x(0)\rangle=0$, $\langle x^2(0)\rangle=\frac{k_BT}{m\omega^2}$, and $\langle x(0)v(0)\rangle=0$. Making Laplace transform of (\ref{ex}) leads to
\begin{eqnarray}
x(s)=\frac{(ms+\xi K(s))x(0)+F(s)+mv(0)}{ms^2+\xi s K(s)+m\omega^2}.
\end{eqnarray}
Then
\begin{eqnarray}
C_x(s)=\frac{\langle x(s)x(0)\rangle}{\langle x^2(0)\rangle}=\frac{ms+\xi K(s)}{ms^2+\xi s K(s)+m\omega^2},
\end{eqnarray}
which results in
\begin{eqnarray}
C_x(t)=1-m\omega^2I(t)
\end{eqnarray}
with $I(s)=\frac{s^{-1}}{ms^2+\xi s K(s)+m\omega^2}$. 

\begin{figure}
  \flushright
  \includegraphics[scale=0.375]{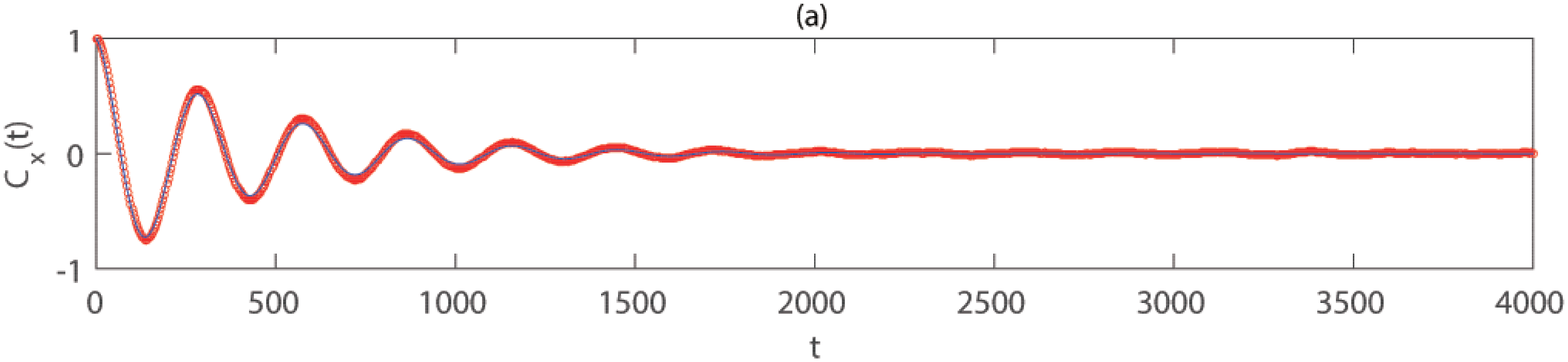}\\
  \includegraphics[scale=0.375]{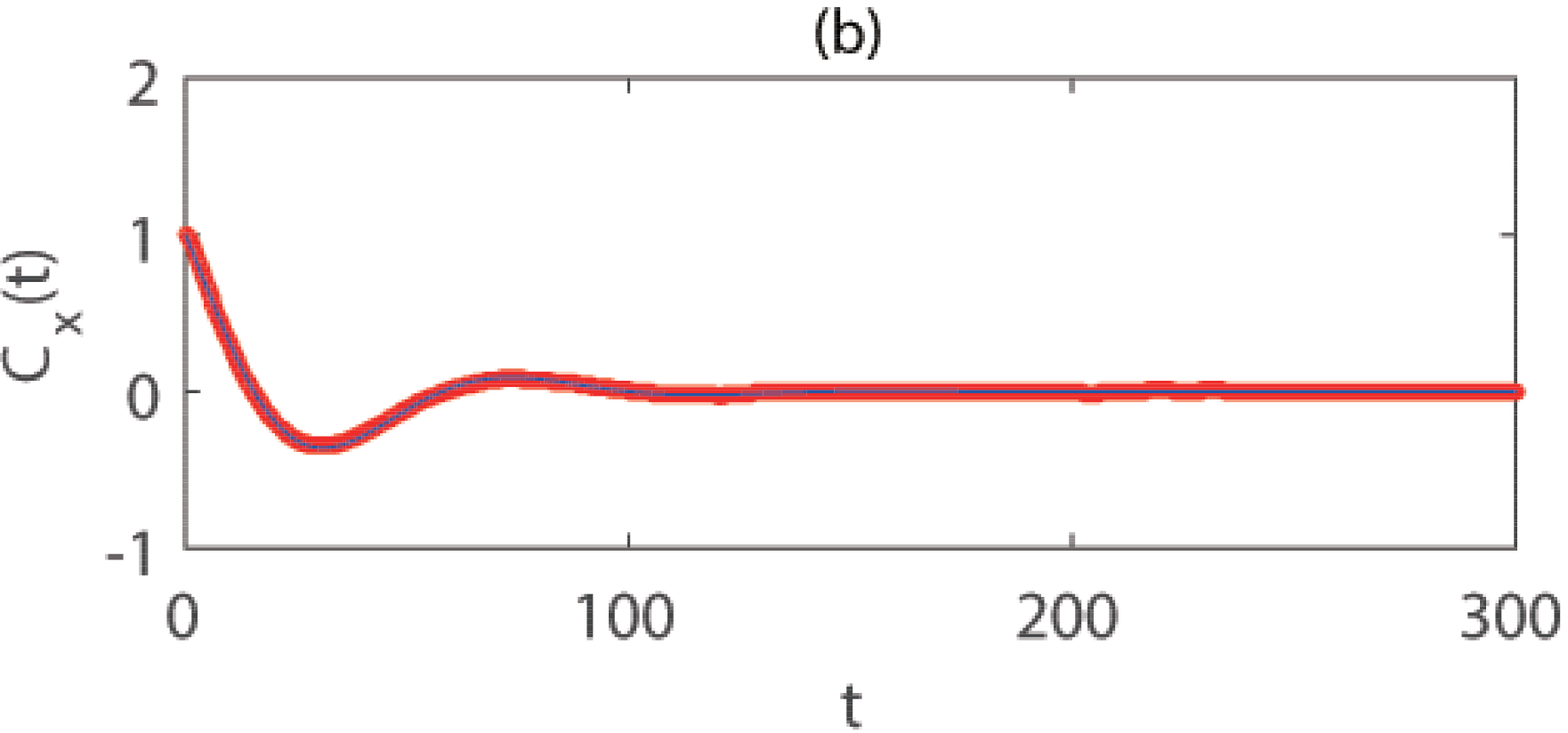}\includegraphics[scale=0.375]{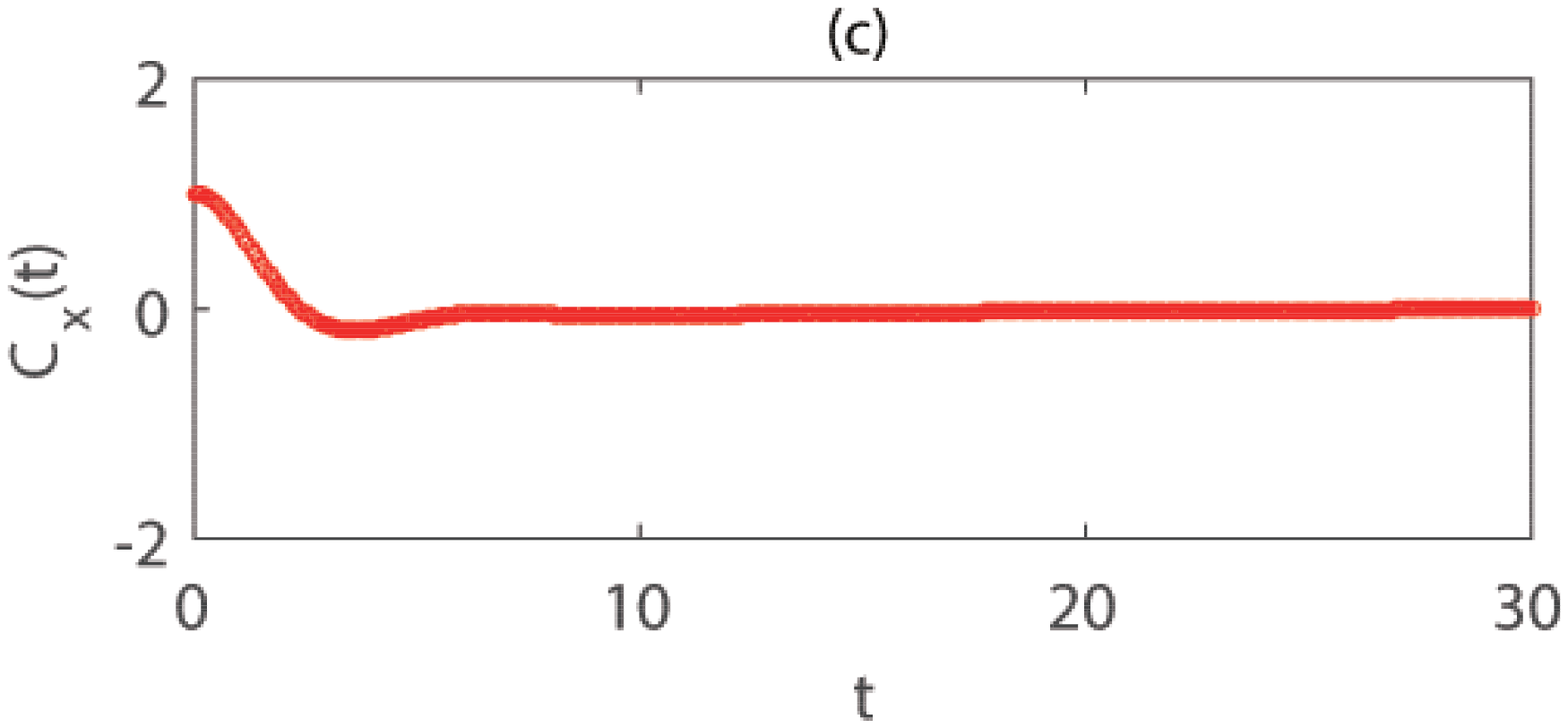}\\
  \includegraphics[scale=0.375]{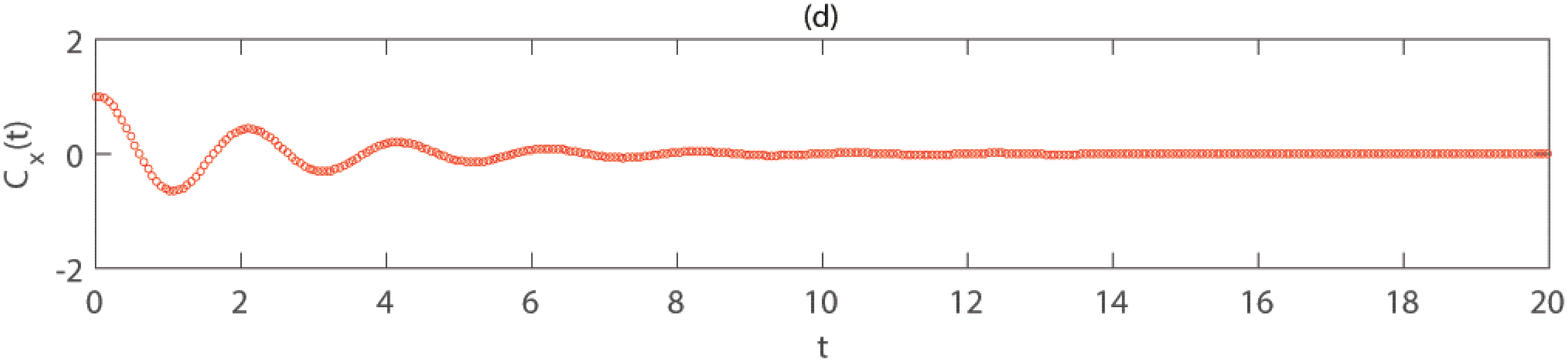}\\
  \caption{Eq. (\ref{cx}) (blue solid line) and the simulation results sampled over $1.5\times10^4$ trajectories (red data points) with $\lambda=0.1$ and $m=1$,  $\omega=0.08$ for (a),   $\omega=0.3$ for (b),  $\omega=0.965$ (just simulation one) for (c), and  $\omega=3$ (just simulation one) for (d). For all the simulations, $H=0.6$. 
   }\label{you}
\end{figure}

\begin{figure}
  \flushright
  \includegraphics[scale=0.38]{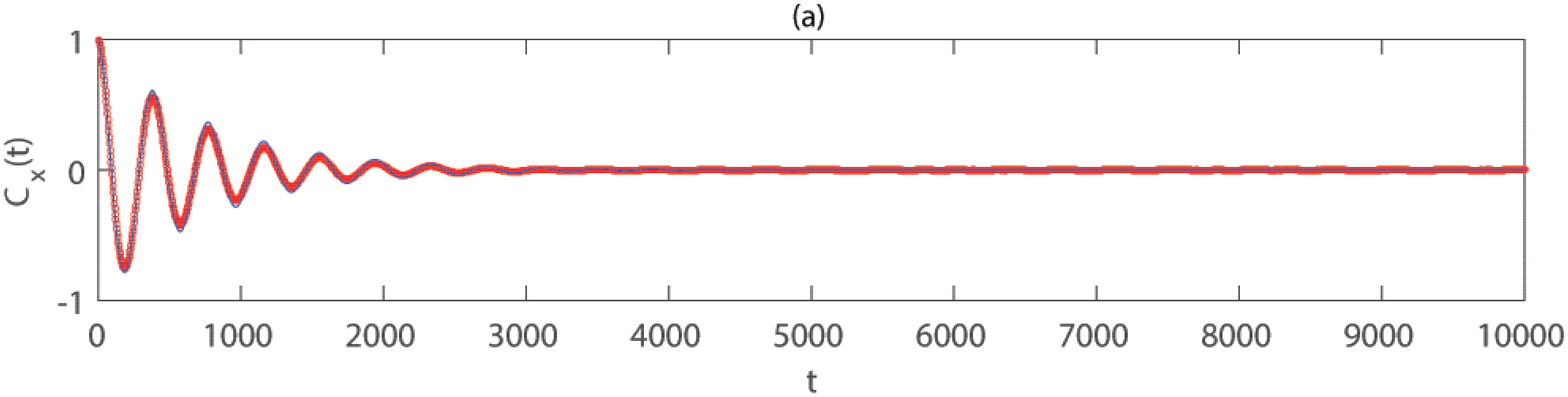}\\
  \includegraphics[scale=0.38]{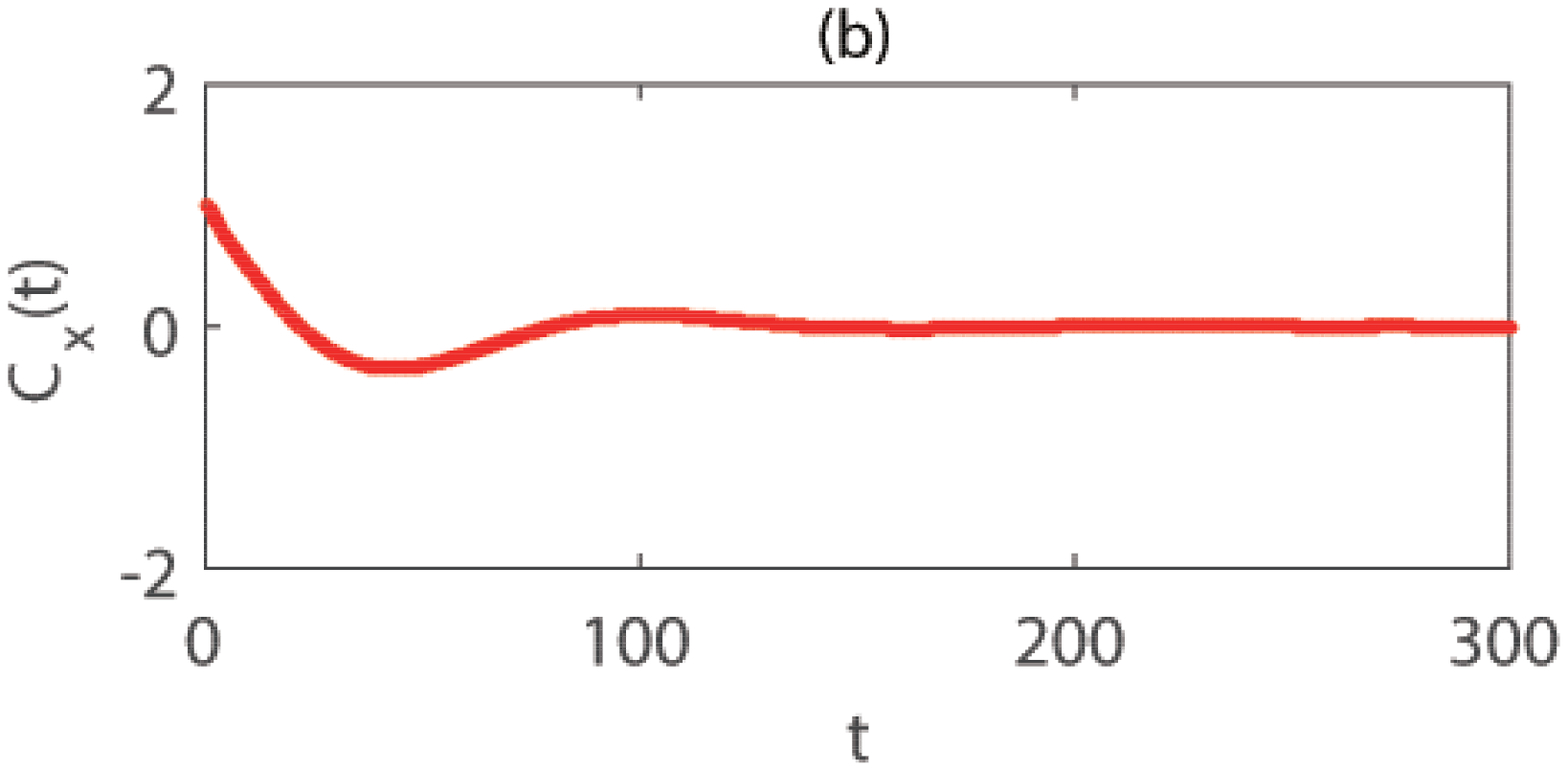}\includegraphics[scale=0.38]{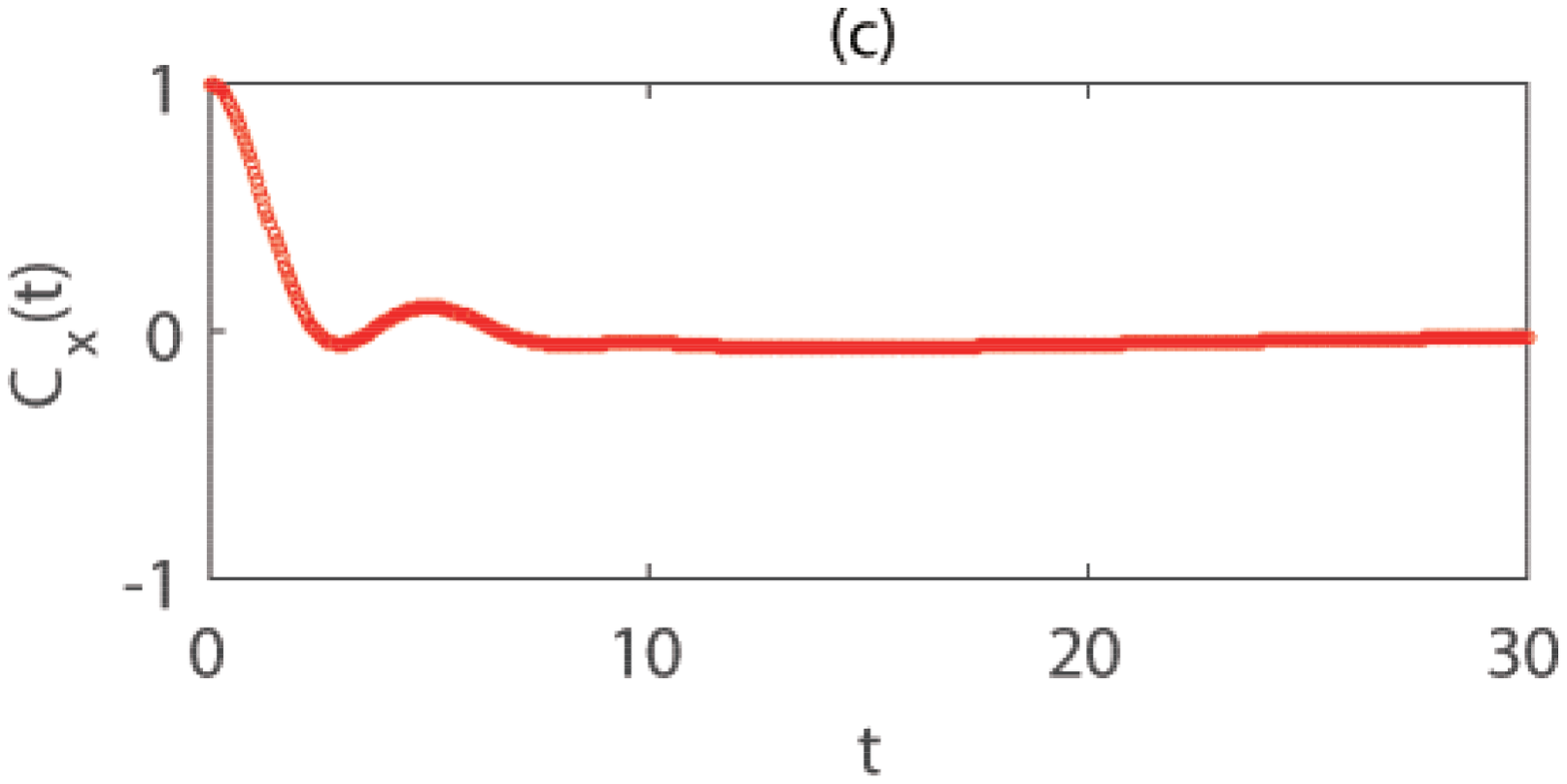}\\
  \includegraphics[scale=0.38]{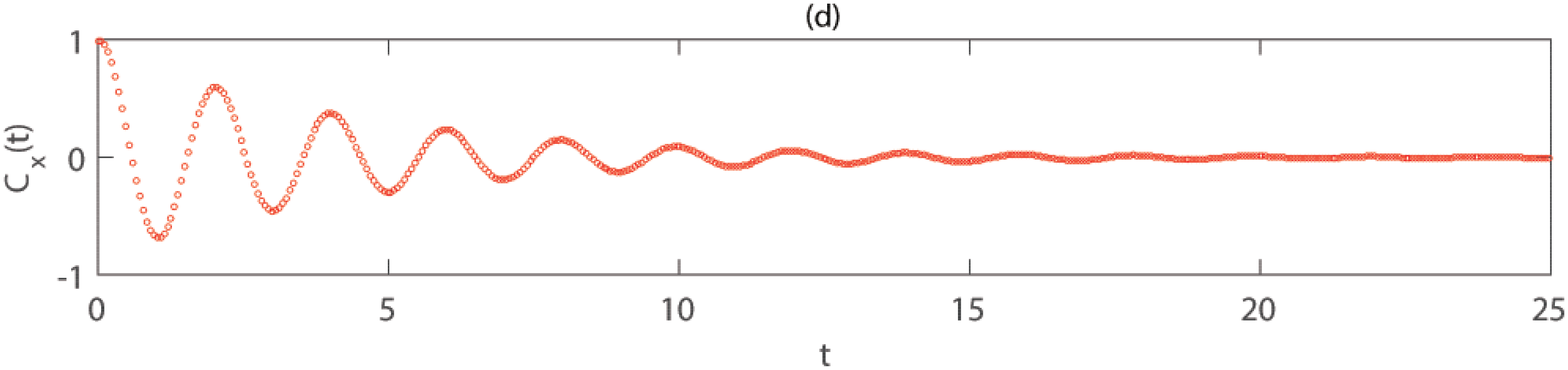}\\
  \caption{
  	Eq. (\ref{cx}) (blue solid line) and the simulation results sampled over $1.5\times10^4$ trajectories (red data points) with $\lambda=0.1$ and $m=1$,  $\omega=0.08$ for (a),   $\omega=0.3$ (just simulation one) for (b),  $\omega=0.965$ (just simulation one) for (c), and  $\omega=3$ (just simulation one) for (d). For all the simulations, $H=0.8$.   	  	
}\label{0-8}
\end{figure}

$K(s)$ and $f(t)$ are given below Eq. (\ref{3.2}). Here we carefully take the asymptotic expression of $f(t)$ as 
\begin{eqnarray}
f(t)\simeq \frac{2 \Gamma(2H)}{(2\lambda)^{2H}}-\frac{2 \Gamma(H+\frac{1}{2})}{(2\lambda)^{H+\frac{1}{2}}}t^{H-\frac{1}{2}}\rme^{-\lambda t}.
\end{eqnarray}
Then
\begin{eqnarray} \label{3.34}
I(s)\simeq\frac{s^{-1}}{as^2-bs^3(s+\lambda)^{-H-\frac{1}{2}}+m\omega^2},
\end{eqnarray}
where $a=m+2\xi \Gamma(2H)(2\lambda)^{-2H}$ and $b=2\xi \Gamma(H+\frac{1}{2})^2(2\lambda)^{-H-\frac{1}{2}}$.
The approximation (\ref{3.34}) is valid just for small $\omega$; for large $\omega$, the convergence region of the approximation of $I(s)$ may be different from the one of $I(s)$. One can note that the approximation in （(\ref{3.34}) is not sensitive to the value of $H$, which is also illustrated by simulations (see Figs. \ref{you} and  \ref{0-8}). We simply take $H=\frac{1}{2}$; for other values of $H$, one can use the techniques in \cite{Burov, Burov-1} to make inverse Laplace transform of $I(s)$, but the calculations are very complicated.
%
%
Then
\begin{eqnarray}\label{i}
I(s)\simeq\frac{s^{-1}(s+\lambda)}{(a-b)s^3+a\lambda s^2+m\omega^2s+m\omega^2\lambda}\simeq\frac{1+\lambda s^{-1}}{a\lambda s^2+m\omega^2s+m\omega^2\lambda}.
\end{eqnarray}
%
Rewrite (\ref{i}) as $a\lambda I(s)=\frac{1}{P(s)}+\frac{s^{-1}\lambda}{P(s)}$ with $P(s)=s^2+\frac{m\omega^2}{a\lambda}s+\frac{m\omega^2}{a}$. Then the solutions of $P(s)=0$ are  $z_1=\alpha+\rmi\beta$ and $z_2=\alpha-\rmi\beta$, where $\alpha=-\frac{m\omega^2}{2a\lambda}$ and $\beta=\frac{\omega\sqrt{4ma\lambda^2-m^2\omega^2}}{2a\lambda}$. The roots are complex on account of the small $\omega$ satisfing $m\omega^2<4a\lambda^2$. Denoting 
\begin{eqnarray}
A_k=\frac{1}{\frac{\rmd P(s)}{\rmd s}|_{s=z_k}},\nonumber
\end{eqnarray}
we have
\begin{eqnarray}
a\lambda I(s)=\sum_{k=1}^{2}\frac{A_k}{s-z_k}+\lambda\sum_{k=1}^{2}\frac{A_ks^{-1}}{s-z_k}
\end{eqnarray}
and
\begin{eqnarray}\label{h}
a\lambda I(t)&=A_1\rme^{z_1t}+A_2\rme^{z_2t}+\lambda\left[\frac{A_1}{z_1}(\rme^{z_1t}-1)+\frac{A_2}{z_2}(\rme^{z_2t}-1)\right]\nonumber\\
&=(A_1+A_2+\lambda\frac{A_1}{z_1}+\lambda\frac{A_2}{z_2})\rme^{\alpha t}\cos(\beta t)\nonumber\\
&\quad +(A_1-A_2+\lambda\frac{A_1}{z_1}-\lambda\frac{A_2}{z_2})\rmi\rme^{\alpha t}\sin(\beta t)-\lambda(\frac{A_1}{z_1}+\frac{A_2}{z_2}),
\end{eqnarray}
where $A_1=\frac{1}{z_1-z_2}$ and $A_2=\frac{1}{z_2-z_1}$.

\begin{figure}
  \flushright
  \quad\,\,\qquad\includegraphics[scale=0.38]{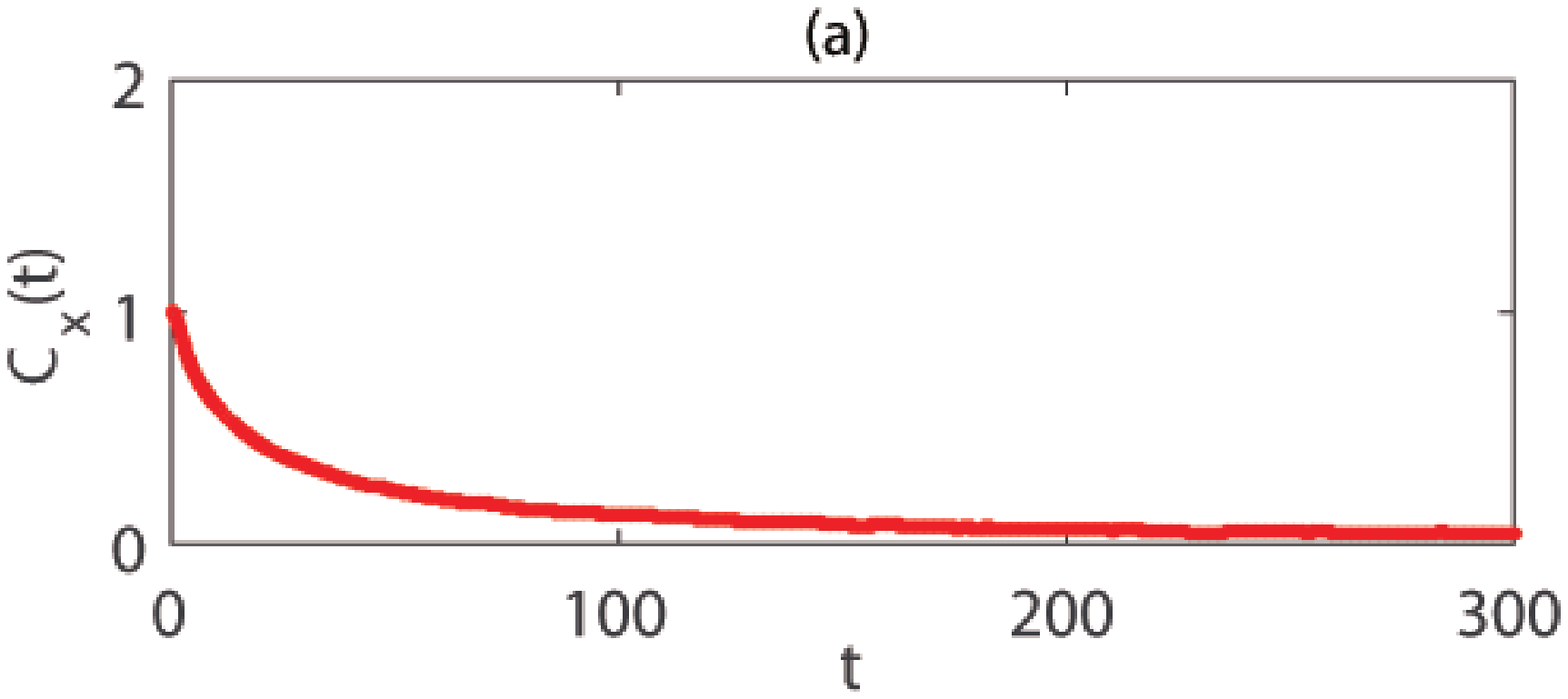}\includegraphics[scale=0.38]{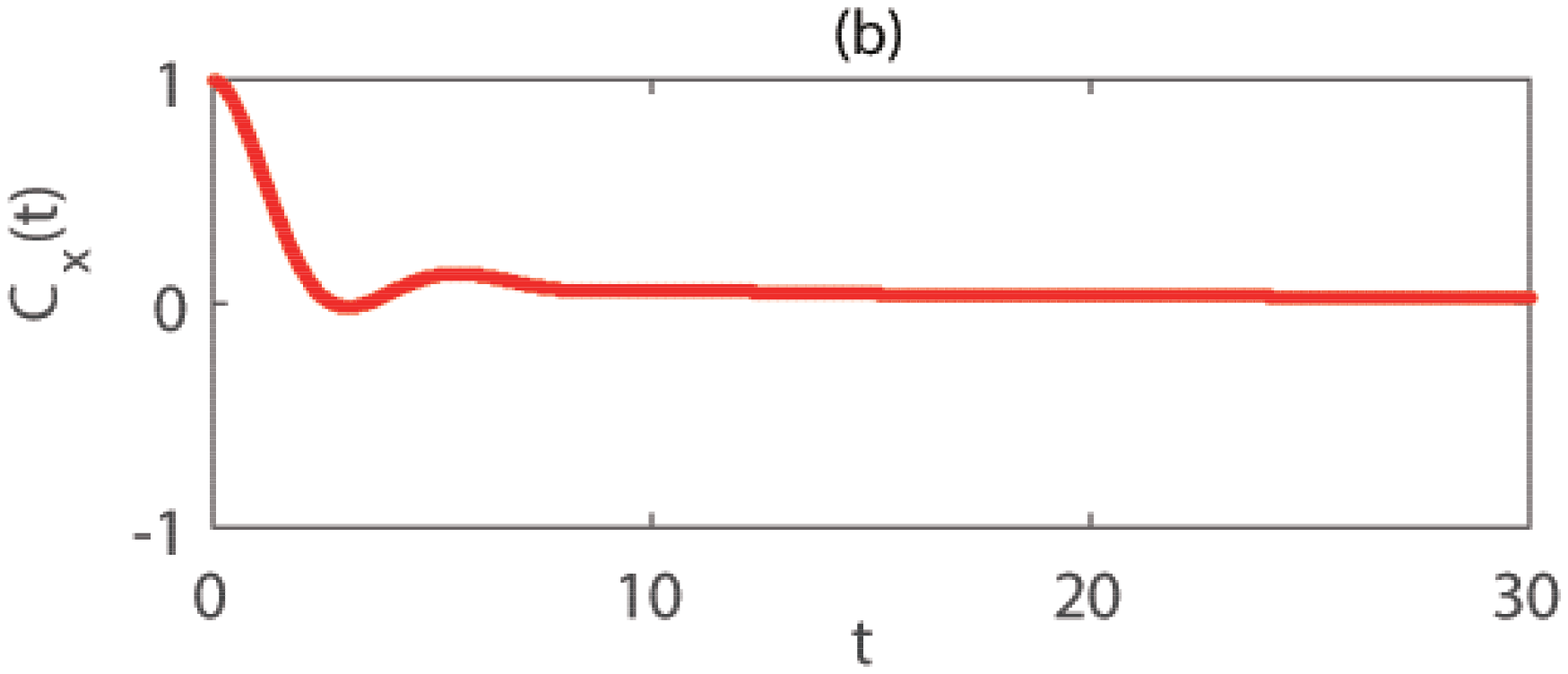}\\
  \quad\quad\includegraphics[scale=0.38]{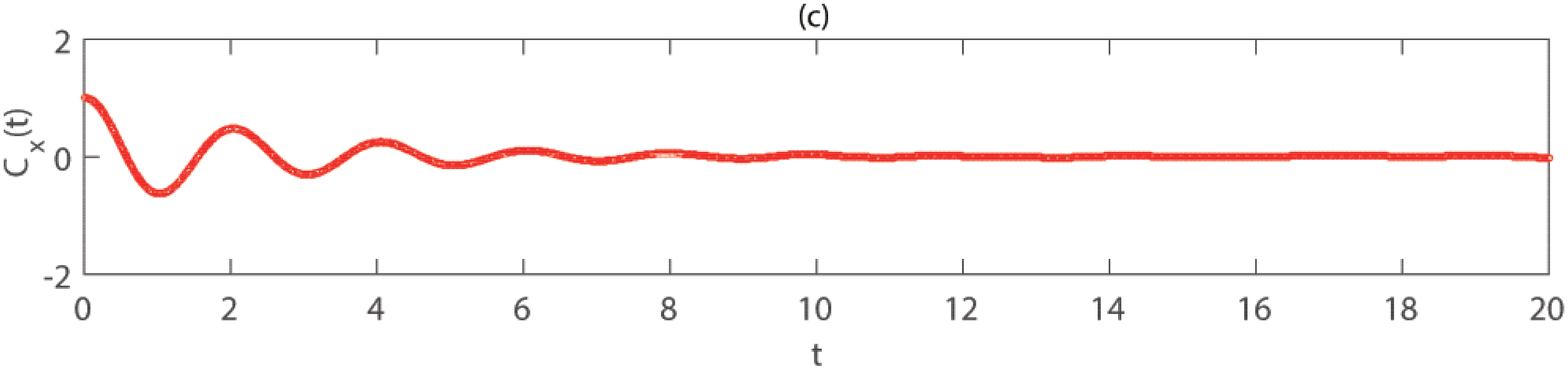}\\
  \caption{Simulation results for $C_x(t)$ with $H=0.625$ and $\lambda=10^{-6}$, sampled over $10^4$
  trajectories. (a): $\omega=0.3$, and the function decays monotonically; (b): $\omega=0.965$, the transition between the motions with and without zero crossing; (c): $\omega=3$, the underdamped regime. It shows that the simulation results are consistent with the theoretical ones in \cite{Burov} since $\lambda$ is very small.}\label{wu}
\end{figure}

Plugging the concrete expressions of $A_k$ into (\ref{h}) leads to
\begin{eqnarray}
I(t)=&\frac{1}{\sqrt{4m\omega^2a\lambda^2-m^2\omega^4}}\rme^{-\frac{m\omega^2}{2a\lambda} t}\sin\left(\frac{\omega\sqrt{4ma\lambda^2-m^2\omega^2}}{2a\lambda} t\right)\nonumber\\
&-\frac{1}{m\omega^2}\rme^{-\frac{m\omega^2}{2a\lambda} t}\cos\left(\frac{\omega\sqrt{4ma\lambda^2-m^2\omega^2}}{2a\lambda} t\right)+\frac{1}{m\omega^2}
\end{eqnarray}
and the normalized displacement correlation function is
\begin{eqnarray}\label{cx}
 C_x(t)&=1-m\omega^2 I(t)\nonumber\\
&=2\lambda \sqrt{\frac{a}{4a\lambda^2-m\omega^2}}\rme^{-\frac{m\omega^2}{2a\lambda} t}\sin\left(\frac{\omega\sqrt{4ma\lambda^2-m^2\omega^2}}{2a\lambda}t+\theta\right),
\end{eqnarray}
where $\theta=\arctan(-\frac{\sqrt{4ma\lambda^2-m^2\omega^2}}{m\omega})$. It shows that for small $\omega$, the phase, amplitude, and period are 
$\theta$, $2\lambda \sqrt{\frac{a}{4a\lambda^2-m\omega^2}}\rme^{-\frac{m\omega^2}{2a\lambda} t}$, and $\frac{4 \pi a\lambda}{\omega\sqrt{4ma\lambda^2-m^2\omega^2}}$,  respectively, being verified by the simulations given in Figure \ref{you} (a) and Figure \ref{you} (b). The simulation results for $H=0.8$ are presented in Figure \ref{0-8}, in which (a) shows that after changing the value of $H$, besides a little bit of the difference of the amplitude, the simulation results are still consistent with Eq. (\ref{cx}). We numerically detect that $C_x(t)$ is always zero crossing. For  $\lambda\rightarrow 0$,  the results for $C_x(t)$ of the fLe \cite{Burov} are recovered; see Figure \ref{wu}.


\section{Conclusion}\label{four}
Tempered fractional Brownian motion is a recently introduced stochastic process, displaying localization diffusion. We further discuss its ergodicity and derive its Fokker-Planck equation. Then we introduce the tempered fractional Langevin equation with the tempered fractional Gaussian noise as internal noise. Both the undamped and overdamped cases are considered, and the evolution of the MSD is carefully investigated, revealing the procedure of transition and finally turning to the ballistic diffusion for long time. The normalized displacement correlation function $C_x(t)$ of the tempered fractional Langevin equation with harmonic potential is explicitly derived for small frequency. 
By the algorithm provided in \ref{AppendixC}, 
almost all of the theoretical results are verified by numerical simulations.

%
%
%

\section*{Acknowledgments}

This work was supported by the National Natural Science Foundation of China under Grant No. 11671182.


\appendix
\setcounter{section}{0}
\section{Derivation of (\ref{qi}) and (\ref{fang})} \label{AppendixA}
For tfBm, denoting $g(z)=C_z^2|z|^{2H}$ with $C_z^2$ defined by (\ref{Cdelta}), the average of $\bar{\delta}^2(x(t))$ is
\begin{eqnarray*}\label{qiwang}
\fl\langle\bar{\delta}^2(x(t))\rangle=\frac{\int_0^{t-\Delta}\langle[x(t'+\Delta)-x(t')]^2\rangle \rmd t'}{t-\Delta} \nonumber \\
\fl\qquad\qquad\,=\frac{\int_0^{t-\Delta}\langle x^2(t'+\Delta)-2x(t'+\Delta)x(t')+x^2(t')\rangle \rmd t'}{t-\Delta} \nonumber \\
\fl\qquad\qquad\,=\frac{\int_0^{t-\Delta}g(t'+\Delta)-[g(t'+\Delta)+C_{t'}^2|t'|^{2H}-C_ {\Delta}^2| \Delta|^{2H}]+C_{t'}^2|t'|^{2H}\rmd t'}{t-\Delta} \nonumber \\
\fl\qquad\qquad\,=C_ {\Delta}^2|\Delta|^{2H},
\end{eqnarray*}
and the variance of $\bar{\delta}^2(x(t))$ is
\begin{eqnarray*}\label{fangcha}
\fl \textrm{Var}[\bar{\delta}^2(x(t))]=\langle [\bar{\delta}^2(x(t))]^2\rangle-\langle \bar{\delta}^2(x(t))\rangle^2 \nonumber \\
\fl=\frac{\int_0^{t-\Delta}\int_0^{t-\Delta}\rmd t_1\rmd t_2\langle[x(t_1+\Delta)-x(t_1)]^2[x(t_2+\Delta)-x(t_2)]^2\rangle}{(t-\Delta)^2}- C_ {\Delta}^4|\Delta|^{4H} \nonumber \\
\fl=\frac{\frac{1}{2}\int_0^{t-\Delta}\int_0^{t-\Delta}\rmd t_1\rmd t_2(g(t_1-t_2+\Delta)+g(t_2-t_1+\Delta)-2g(t_1-t_2))}{(t-\Delta)^2} \nonumber \\
\fl\simeq\frac{\frac{1}{2}\int_0^{t-\Delta}\int_0^{t-\Delta}\rmd t_1\rmd t_2(-G(t_1-t_2+\Delta)-G(t_2-t_1+\Delta)+2G(t_1-t_2))^2}{(t-\Delta)^2}
\nonumber\\
\fl=\frac{\int_0^{t-\Delta}\int_{t'}^{t-\Delta}\rmd t_1\rmd t'(-G(\Delta+t')-G(\Delta-t')+2G(t'))^2}{(t-\Delta)^2}\nonumber\\
\fl=\frac{\int_0^{t-\Delta}\rmd t'(-G(\Delta+t')-G(\Delta-t')+2G(t'))^2(t-t'-\Delta)}{(t-\Delta)^2}\nonumber\\
\fl=\frac{4\Gamma^2(H+\frac{1}{2})}{(2\lambda)^{2H+1}}\frac{\int_0^{\frac{t}{\Delta}-1}\rmd\tau\Delta^{2H}(t-\Delta-\Delta\tau)Q(\tau)}{(t-\Delta)^2} \nonumber\\
\fl=\frac{4\Gamma^2(H+\frac{1}{2})\Delta^{2H}}{(2\lambda)^{2H+1}(t-\Delta)}\int_0^{\frac{t}{\Delta}-1}Q(\tau)\rmd\tau -\frac{4\Gamma^2(H+\frac{1}{2})\Delta^{2H+1}}{(2\lambda)^{2H+1}(t-\Delta)^2}\int_0^{\frac{t}{\Delta}-1}\tau Q(\tau)\rmd\tau\nonumber\\
\fl\simeq D\frac{4\Gamma^2(H+\frac{1}{2})}{(2\lambda)^{2H+1}}\Delta^{2H}t^{-1}
\end{eqnarray*}
for long times and moderate $\lambda$, 
where
\begin{eqnarray*}
G(t)=\frac{2\Gamma(H+\frac{1}{2})}{(2\lambda)^{H+\frac{1}{2}}}|t|^{H-\frac{1}{2}}\rme^{-\lambda |t|},
\end{eqnarray*}
\begin{eqnarray*}
Q(t)=\left[(1+t)^{H-\frac{1}{2}}\rme^{-\lambda\Delta(1+t)}+
|t-1|^{H-\frac{1}{2}}\rme^{-\lambda|\Delta(t-1)|}-2t^{H-\frac{1}{2}}\rme^{-\lambda\Delta t}\right]^2,
\end{eqnarray*}
\begin{eqnarray*}
D=\int_0^{\infty}\rmd\tau[(1+\tau)^{H-\frac{1}{2}}\rme^{-\lambda\Delta(1+\tau)}+
|\tau-1|^{H-\frac{1}{2}}\rme^{-\lambda|\Delta(\tau-1)|}-2\tau^{H-\frac{1}{2}}\rme^{-\lambda\Delta\tau}]^2.
\end{eqnarray*}

\section{Asymptotic behavior of the covariance function of tfGn} \label{AppendixB}
\begin{figure}[!htb]
	\flushright
	\,\,\,\includegraphics[scale=0.4]{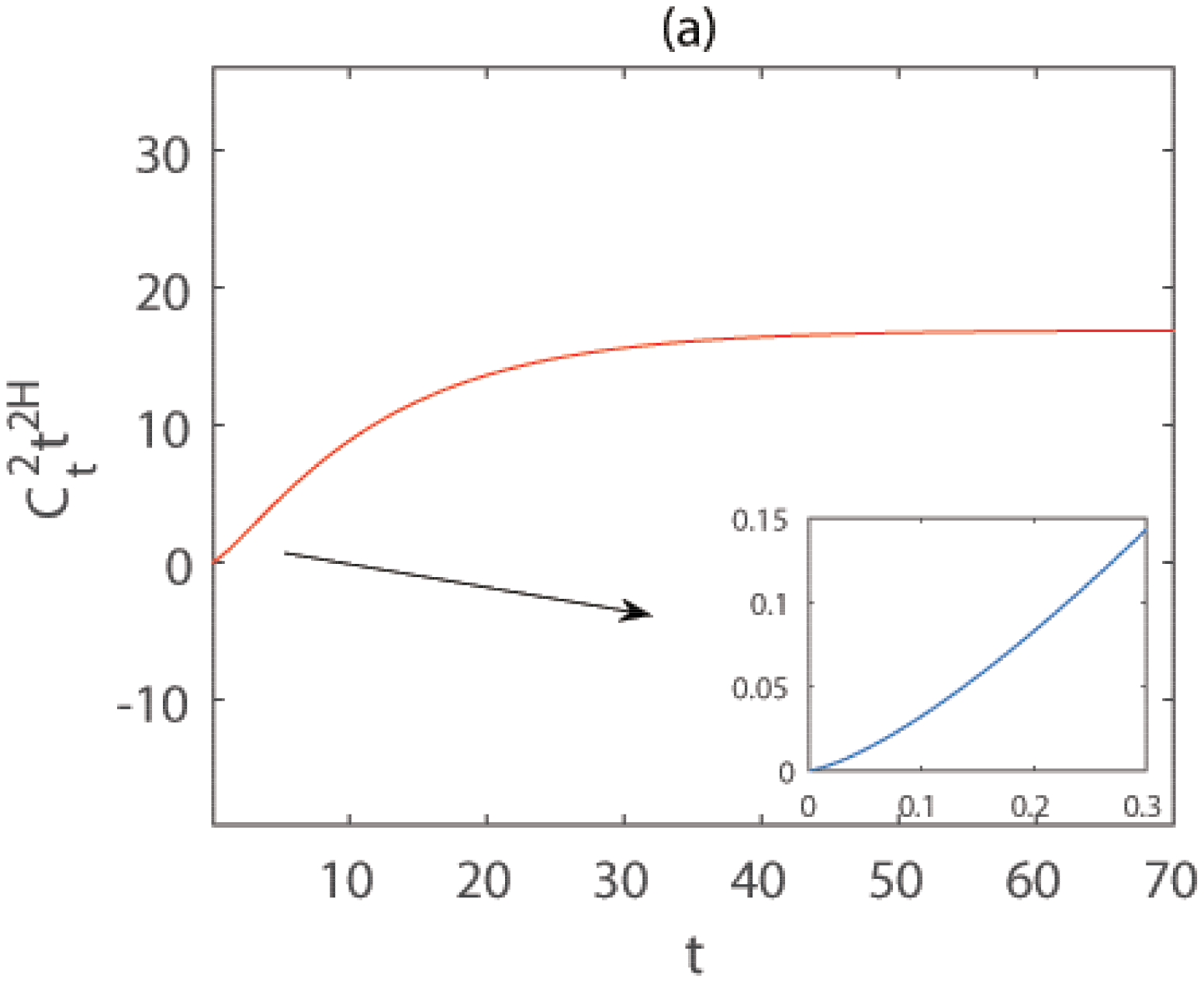}\includegraphics[scale=0.4]{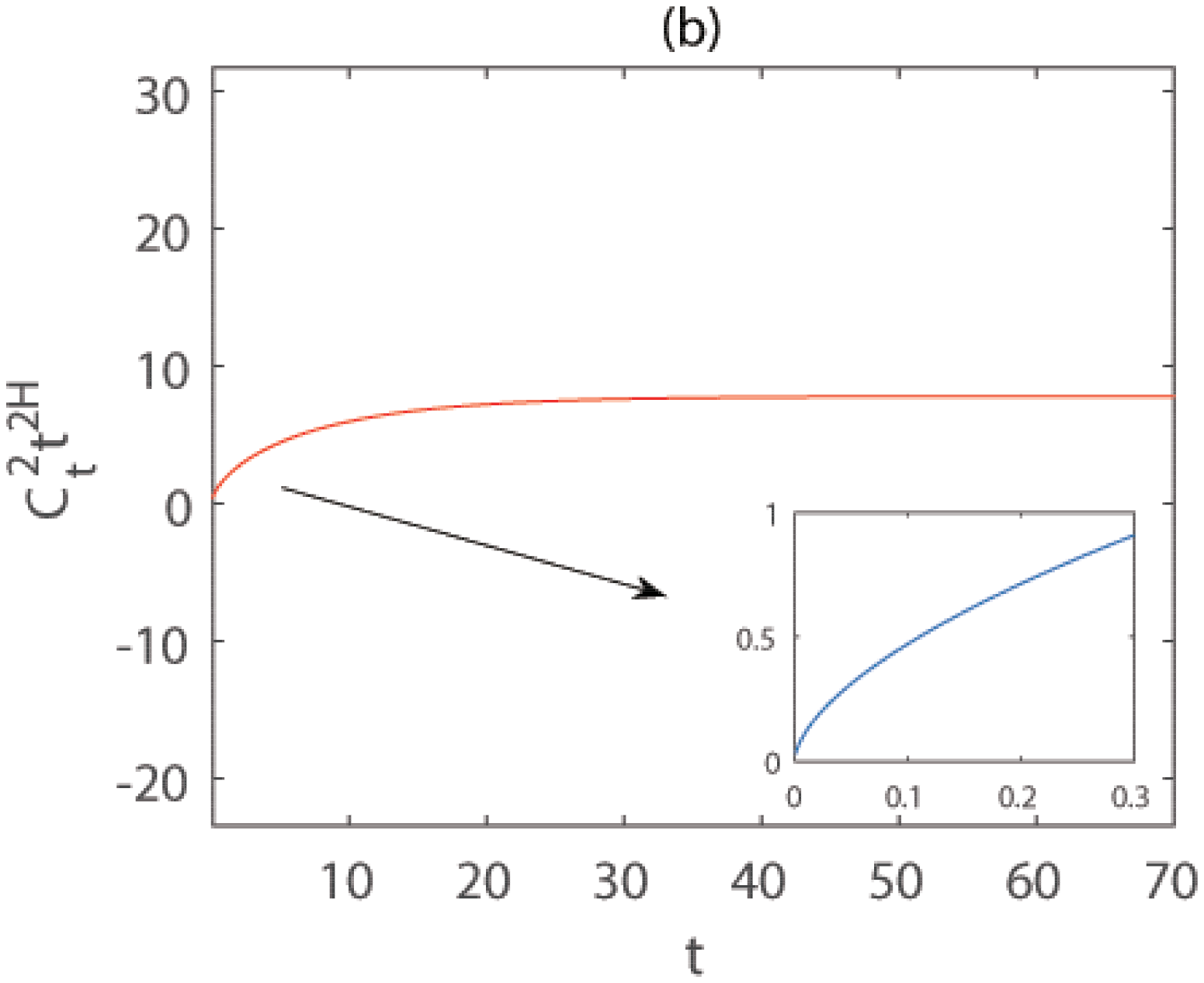}\\
	\caption{Evolution of the asymptotic second integration of $K(t)$ defined in Eq. (\ref{Kt1}) and Eq. (\ref{LV}). (a): $H=0.7$ and $\lambda=0.1$. It is clear that $C_t^2t^{2H}$ is a convex function when $t$ is small and then along with $t$'s increasing, $C_t^2t^{2H}$ becomes a concave function. So the second derivation of $C_t^2t^{2H}$ goes from positive to negative, and then it approaches to zero. (b): $H=0.3$ and $\lambda=0.1$. It is clear that $C_t^2t^{2H}$ is always a concave function. So the second derivation of $C_t^2t^{2H}$ is always negative, and in the end it tends to zero.}\label{5}
\end{figure}

Fixing the value of $\lambda$ and considering sufficiently small $t$, 
we have
\begin{eqnarray} \label{Kt1}
K(t)&=2\langle\gamma(0)\gamma(t)\rangle\nonumber\\
&=\frac{1}{h^2}(C_{t+h}^2|t+h|^{2H}+C_{t-h}^2|t-h|^{2H}-2C_t^2|t|^{2H})\\
&\simeq C_t^2\frac{1}{h^2}(|t+h|^{2H}+|t-h|^{2H}-2|t|^{2H})\nonumber\\
&\simeq 2C_t^2H(2H-1)t^{2H-2},\nonumber
\end{eqnarray}
where $C_t^2$ is a constant. 
Figure \ref{5} is plotting the asymptotic second integration of $K(t)$, i.e., $C_t^2|t|^{2H}$. The convexity (concavity) of $C_t^2|t|^{2H}$ implies the positivity (negativity) of $K(t)$.
It is clear that  $K(t)>0$ (see Figure \ref{5}(a)) for $\frac{1}{2}<H<1$, and  $K(t)<0$ (see Figure \ref{5}(b)) for $0<H<\frac{1}{2}$.

Fixing $\lambda$ and letting $t \rightarrow +\infty$, i.e., $\lambda t$ is large, we have
\begin{eqnarray}\label{LV}
\fl K(t)=2\langle\gamma(0)\gamma(t)\rangle\nonumber\\
\fl~~~~~~=\frac{1}{h^2}\left(C_{t+h}^2|t+h|^{2H}+C_{t-h}^2|t-h|^{2H}-2C_t^2|t|^{2H}\right)\nonumber\\
\fl~~~~~~= \frac{1}{h^2}\left[(C_{t+h}^2+C_t^2-C_t^2)|t+h|^{2H}+(C_{t-h}^2+C_t^2-C_t^2)|t-h|^{2H}-2C_t^2|t|^{2H}\right]\nonumber\\
\fl~~~~~~=\frac{C_t^2}{h^2}|t|^{2H}\left(\left|1+\frac{h}{t}\right|^{2H}+\left|1-\frac{h}{t}\right|^{2H}-2\right)\\
\fl ~~~~~~~~~+\frac{1}{h^2}\left[(C_{t+h}^2-C_t^2)|t+h|^{2H}+(C_{t-h}^2-C_t^2)|t-h|^{2H}\right].\nonumber
\end{eqnarray}
Using Taylor's series expansion, along with $C_t^2=\frac{2\Gamma(2H)}{(2\lambda|t|)^{2H}}-\frac{2\Gamma(H+\frac{1}{2})K_H(\lambda|t|)}{\sqrt{\pi}(2\lambda|t|)^H}=\bar{A}|t|^{-2H}-\bar{B}|t|^{-H}K_H(\lambda|t|)$ and  $K_H(\lambda t)\simeq\sqrt{\pi}(2\lambda t)^{-\frac{1}{2}}\rme^{-\lambda t}$ as $\lambda t$ is large, the first term in (\ref{LV}) is
$
2C_t^2H(2H-1)t^{2H-2}\simeq2\bar{A}H(2H-1)t^{-2}-2\bar{B}H(2H-1)\sqrt{\pi}(2\lambda)^{-\frac{1}{2}}t^{H-\frac{5}{2}}\rme^{-\lambda t},
$
and the second term in (\ref{LV}) is
\begin{eqnarray*}
\fl\frac{1}{h^2}\left[(C_{t+h}^2-C_t^2)|t+h|^{2H}+(C_{t-h}^2-C_t^2)|t-h|^{2H}\right]\\
\fl=-\frac{\bar{A}}{h^2}\left\{|1+\frac{h}{t}|^{2H}+|1-\frac{h}{t}|^{2H}-2\right\}\\
\fl ~~+\frac{\bar{B}}{h^2}\left\{-|t+h|^HK_H(\lambda|t+h|)+\frac{|t+h|^{2H}}{t^H}K_H(\lambda t)-|t-h|^HK_H(\lambda|t-h|) \right.
\\
\fl ~~ \left. +\frac{|t-h|^{2H}}{t^H}K_H(\lambda t)\right\}\\
\fl\simeq-2\bar{A}H(2H-1)t^{-2}
\\
\fl~~ -\sqrt{\pi}(2\lambda)^{-\frac{1}{2}}\frac{\bar{B}}{h^2}\left\{g(t+h)+g(t-h)-\left[(1+\frac{h}{t})^{2H}+(1-\frac{h}{t})^{2H}\right]g(t)\right\}\\
\fl\simeq-2\bar{A}H(2H-1)t^{-2}
 -\sqrt{\pi}(2\lambda)^{-\frac{1}{2}}\frac{\bar{B}}{h^2}\left\{g(t+h)+g(t-h)-2g(t)\right\}\\
\fl\simeq-2\bar{A}H(2H-1)t^{-2}-\sqrt{\pi}(2\lambda)^{-\frac{1}{2}}\bar{B}\frac{\rmd^2g(t)}{\rmd t^2}\\
\fl\simeq-2\bar{A}H(2H-1)t^{-2}-\sqrt{\pi}(2\lambda)^{-\frac{1}{2}}\bar{B}\lambda^2t^{H-\frac{1}{2}}\rme^{-\lambda t},
\end{eqnarray*}
where $g(x)=x^{H-\frac{1}{2}}\rme^{-\lambda x}$. Combining the above two estimations leads to
\begin{eqnarray}
&K(t)\simeq-\frac{2\Gamma(H+\frac{1}{2})\lambda^2}{(2\lambda)^{H+\frac{1}{2}}}t^{H-\frac{1}{2}}\rme^{-\lambda t}<0
\end{eqnarray}
for large $t$; see Figure \ref{5}. The $K(t)$ can also be obtained by making second derivative on the asymptotic expression of $C_t^2t^{2H}$ for large $t$.
So the asymptotic behavior of tfGn's covariance function is $\langle\gamma(0)\gamma(t)\rangle\simeq-\frac{\Gamma(H+\frac{1}{2})\lambda^2}{(2\lambda)^{H+\frac{1}{2}}}t^{H-\frac{1}{2}}\rme^{-\lambda t}$ for large $ t$ and $\langle\gamma(0)\gamma(t)\rangle\simeq C_t^2H(2H-1)t^{2H-2}$ with a constant $C_t^2$ for small $ t$.

\section{Algorithm for numerical simulations} \label{AppendixC}
To generate tempered fractional Gaussian noise $X_0,X_1,\cdots$, we adopt the Hosking method \cite{Dieker}, which works for the general stationary Gaussian process.
 The key observation of this algorithm is to generate $X_{n+1}$ by the conditional distribution of $X_{n+1}$ given $X_n,\cdots,X_0$ recursively.
The covariance function of tfGn is
\begin{equation*}
  \rho(k):=\mathbb{E}X_nX_{n+k},
\end{equation*}
for $n,k=0,1,2,\cdots$. Note that $\langle B_{\alpha,\lambda}^2(t)\rangle=C_t^2|t|^{2H}$ from (\ref{covariance}) and $B_{\alpha,\lambda}(t)$ has the  stationary increments, $X_n:=B_{\alpha,\lambda}(n+1)-B_{\alpha,\lambda}(n)\sim N(0,C_1^2)$, which means $\rho(0)=\mathbb{E}(X_n^2)=C_1^2$. Furthermore, let $D(n)=\left(\rho(i-j)\right)_{i,j=0,\cdots,n}$
be the covariance matrix and $c(n)$ be the $(n+1)-$column vector with elements $c(n)_k=\rho(k+1),k=0,\cdots,n$. Define the $(n+1)\times(n+1)$ flipping matrix $F(n)=(\mathbf{1}(i=n-j))_{i,j=0,\cdots,n}$, where $\mathbf{1}$ denotes the indicator function.

We claim that the conditional distribution of $X_{n+1}$ is Gaussian with expectation $\mu_n$ and variance $\sigma_n^2$ given by
\begin{equation}\label{C:1}
  \mu_n:=c(n)'D(n)^{-1}
  \left(\begin{array}{c} X_n \\ \vdots \\ X_1 \\ X_0 \end{array}\right),
  \quad \sigma_n^2:=C_1^2-c(n)'D(n)^{-1}c(n).
\end{equation}
To avoid matrix inversion in (\ref{C:1}) in each step, 
define $d(n):=D(n)^{-1}c(n)$, and $\tau_n:=d(n)'F(n)c(n)=c(n)'F(n)d(n)$.
Split the matrix $D(n+1)$ as follows:
\begin{eqnarray*}
D(n+1)&=
\left(\begin{array}{cc} C_1^2 & c(n)' \\ c(n) & D(n) \end{array} \right)\\
&=\left(\begin{array}{cc} D(n) & F(n)c(n) \\ c(n)'F(n) & C_1^2 \end{array} \right).
\end{eqnarray*}
With some simple calculations, one gets
\begin{eqnarray}\label{C:2}
  D(n+1)^{-1}&=
  \frac{1}{\sigma_n^2}\left(
\begin{array}{cc}
  1 & -d(n)' \\ -d(n) & \sigma_n^2D(n)^{-1}+d(n)d(n)'
\end{array}
\right)\\\label{C:3}
&=
\frac{1}{\sigma_n^2}\left(
\begin{array}{cc}
  \sigma_n^2D(n)^{-1}+F(n)d(n)d(n)'F(n) & -F(n)d(n) \\ -d(n)'F(n) & 1
\end{array}
\right).
\end{eqnarray}
From (\ref{C:2}), for each $x\in\mathbb{R}^{n+1}$ and $y\in\mathbb{R}$, we have
\begin{equation*}
  \left(\begin{array}{cc} y & x' \end{array}\right)
  D(n+1)^{-1} \left(\begin{array}{c} y \\ x \end{array}\right)
  =\frac{(y-d(n)'x)^2}{\sigma_n^2}+x'D(n)^{-1}x.
\end{equation*}
This implies that the conditional distribution of $X_{n+1}$ is indeed Gaussian with expectation $\mu_n$ and variance $\sigma_n^2$. On the other hand, by (\ref{C:3}), some recursions are as follows:
\begin{equation*}
  \sigma_{n+1}^2=\sigma_n^2-\frac{\left(\rho(n+2)-\tau_n\right)^2}{\sigma_n^2}
\end{equation*}
and
\begin{equation*}
  d(n+1)=\left(\begin{array}{c}
    d(n)-\phi_nF(n)d(n) \\ \phi_n
  \end{array}\right)
\end{equation*}
with
\begin{equation*}
  \phi_n=\frac{\rho(n+2)-\tau_n}{\sigma_n^2}.
\end{equation*}
We start the recursion with $\mu_0=\rho(1)X_0,~\sigma_0^2=C_1^2-\rho(1)^2$ and $\tau_0=\frac{1}{C_1^2}\rho(1)^2$. Taking cumulative sums on the generated tfGn samples $X_0,\cdots,X_n$, one obtains the tempered fractional Brownian motion sample $B_{\alpha,\lambda}(k),k=0,\cdots,n$.


Next, we consider to generate a series tfGn samples of number $N$ for obtaining $B_{\alpha,\lambda}(T)$. The scaling
property of tfBm shows that $B_{\alpha,\lambda}(T)=\left(\frac{T}{N}\right)^H B_{\alpha,\lambda T/N}(N)$ \cite{Meerschaert:00}. Denoting  $\gamma(T):=\gamma_{\alpha,\lambda}(T)=\left(\frac{T}{N}\right)^{H-1}\gamma_{\alpha,\lambda T/N}(N)$, i.e., $h=\frac{T}{N}$ in (\ref{2.9}), 
for simulating the second moment of position $\langle x^2(t)\rangle$, we solve (\ref{PDC}) with the scheme:
\begin{eqnarray*}
\fl m~\frac{v(t_{n+1})-v(t_n)}{h}\\
\fl=-\xi\int_0^{t_{n+1}}K(t_{n+1}-\tau)v(\tau)\rmd\tau+\sqrt{2k_BT\xi}~\gamma(t_{n+1})\\
\fl=-\xi\left(\frac{h}{2}\left(K(t_{n+1})v(0)+K(0)v(t_{n+1})\right)+h\sum_{i=1}^nK(t_i)v(t_{n+1-i})\right)
  +\sqrt{2k_BT\xi}~\gamma(t_{n+1}),
\end{eqnarray*}
where $h$ is the step length, and $K(t)=\frac{1}{h^2}(C_{t+h}^2|t+h|^{2H}+C_{t-h}^2|t-h|^{2H}-2C_t^2|t|^{2H})$.
In simulating the normalized displacement correlation function $C_x(t)$, we take the initial conditions as $v_0\sim N(0,\frac{k_BT}{m})$, and $x_0\sim U[-r,r]$, where $r=\sqrt{\frac{3k_BT}{m}}\frac{1}{\omega}$.


\section*{References}
\begin{thebibliography}{<num>}
\bibitem{Shlesinger} M F Shlesinger, G M Zaslavsky and U Frisch 1994 {\it L\'{e}vy Flights and Related Topics in Physics} (France: Springer Verlag).

\bibitem{Hughes} B D Hughes 1995 {\it Random Walks and Random Environments} (Oxford: Oxford Science)
\bibitem{Metzler} R Metzler and J Klafter 2000 {\it Phys. Rep.} {\bf 339} 1
\bibitem{Drysdale} P M Drysdale and P A Robinson 1998 {\it Phys. Rev. E} {\bf 58} 5382
\bibitem{Metzler:01} R Metzler and J Klafter 2004 {\it J. Phys. A} {\bf 37} R161
\bibitem{Schumer:03} R Schumer, M M Meerschaert and B Baeumer 2009 {\it J. Geophys. Res.} {\bf 114} F00A07
\bibitem{Bruno:00} R Bruno, L Sorriso-Valvo, V Carbone and B Bavassano 2004 {\it Europhys. Lett.} {\bf 66} 146

\bibitem{Kou} S C Kou and X Sunney Xie 2004 {\it Phys. Rev. Lett.} {\bf 93} 180603

\bibitem{Pipiras} V Pipiras and M Taqqu 2000 
{\it Probab. Theory Related Fields} {\bf 118} 251
\bibitem{Deng} W H Deng and E Barkai 2009 {\it Phys. Rev. E} {\bf 79} 011112
\bibitem{Meerschaert:02} Mark M Meerschaert and Alla Sikorskii 2011 {\it  Stochastic Models for Fractional Calculus} (De Gruyter)
\bibitem{Eric} Eric Lutz 2001 {\it Phys. Rev. E} {\bf 64} 051106


\bibitem{Coffey:04} W T Coffey, Yu P Kalmykov and J T Waldron 2004 {\it  The Langevin
Equation} (World Scientific, Singapore)

\bibitem{Mandelbrot:68} B. B. Mandelbrot and J. W. van Ness 1968 {\it SIAM Rev.} {\bf 10} 422




\bibitem{Meerschaert:00} M M Meerschaert and F Sabzikar 2013 {\it Statist. Probab. Lett.} {\bf 83} 2269
\bibitem{Meerschaert:01} M M Meerschaert and F Sabzikar 2014 
{\it Stoch. Process. Appl.} {\bf 124} 2363

\bibitem{Kubo} R Kubo 1966 {\it Rep. Prog. Phys.} {\bf 29} 255

\bibitem{Srivastava} H M Srivastava 1979 {\it Publ. Inst. Math.} {\bf 26} 273

\bibitem{Podlubny} I Podlubny 1999 {\it Fractional Differential Equations} (London: Academic Press)

\bibitem{Burov} S Burov and E Barkai 2008 {\it Phys. Rev. E} {\bf 78} 031112

\bibitem{Burov-1} S Burov and E Barkai 2008 {\it Phys. Rev. Lett.} {\bf 100} 070601
\bibitem{Dieker} T Dicker 2002 {\it Simulation of Fractional Brownian Motion} (Master's thesis, University of Twente)



\end{thebibliography}
\end{document}